\documentclass[acmlarge,screen,nonacm]{acmart}

\usepackage{graphicx}
\usepackage{subcaption}
\usepackage{xcolor}
\usepackage{soul}
\usepackage{afterpage}
\usepackage{makecell}
\sethlcolor{yellow} % default, but we will override locally
\usepackage{float}

\usepackage{listings}
\usepackage{array}
\usepackage{multirow}
\AtBeginDocument{%
  }

\begin{document}

\title{CHOMP: Multimodal Chewing-Side Detection with Earphones}

\author{Jonas Hummel}
\authornote{Both authors contributed equally to this research.}
\email{jonas.hummel@kit.edu}
\orcid{0009-0005-8563-6175}
\author{Maximilian Burzer}
\authornotemark[1]
\email{maximilian.burzer@kit.edu}
\orcid{0009-0000-9628-8667}
\affiliation{%
  \institution{Karlsruhe Institute of Technology}
  \city{Karlsruhe}
  \country{Germany}
}

\author{Felix Schlotter}
\email{upsxl@student.kit.edu}
\orcid{0009-0001-6148-0891}
\affiliation{%
  \institution{Karlsruhe Institute of Technology}
  \city{Karlsruhe}
  \country{Germany}
}

\author{Michael Küttner}
\email{michael.kuettner@kit.edu}
\orcid{0009-0000-9021-0359}
\affiliation{%
  \institution{Karlsruhe Institute of Technology}
  \city{Karlsruhe}
  \country{Germany}
}

\author{Tobias King}
\email{tobias.king@kit.edu}
\orcid{0009-0006-7289-3356}
\affiliation{%
  \institution{Karlsruhe Institute of Technology}
  \city{Karlsruhe}
  \country{Germany}
}

\author{Qiang Yang}
\email{qiang.yang@cl.cam.ac.uk}
\orcid{0000-0002-5202-7892}
\affiliation{%
  \institution{University of Cambridge}
  \city{Cambridge}
  \country{United Kingdom}
}

\author{Katharina Scherf}
\email{k.scherf.leibniz-lsb@tum.de}
\orcid{0000-0001-8315-5400}
\affiliation{%
  \institution{Technical University of Munich}
  \city{Munich}
  \country{Germany}
}

\author{Cecilia Mascolo}
\email{cm542@cam.ac.uk}
\orcid{0000-0001-9614-4380}
\affiliation{%
  \institution{University of Cambridge}
  \city{Cambridge}
  \country{United Kingdom}
}

\author{Michael Beigl}
\email{michael.beigl@kit.edu}
\orcid{0000-0001-5009-2327}
\affiliation{%
  \institution{Karlsruhe Institute of Technology}
  \city{Karlsruhe}
  \country{Germany}
}

\author{Tobias R{\"o}ddiger}
\email{tobias.roeddiger@ipai-foundation.ai}
\orcid{0000-0002-4718-9280}
\affiliation{%
  \institution{Innovation Park Artificial Intelligence}
  \city{Heilbronn}
  \country{Germany}
}

\renewcommand{\shortauthors}{Jonas Hummel \& Maximilian Burzer et al.}

\begin{abstract}

Chewing-side preference (CSP) is a risk factor for temporomandibular disorders (TMDs) and a behavioral manifestation. Although TMDs affect roughly one-third of the global population, assessment relies on clinical examinations and self-reports, providing limited insight into everyday jaw function. We present CHOMP, the first earphone-based chewing-side detection system for continuous CSP monitoring. Using OpenEarable 2.0, we collected multimodal data from 20 participants with microphones, a bone-conduction microphone, IMU, PPG, and a pressure sensor across diverse foods, activities, and acoustic-interference conditions. CHOMP models paired-ear temporal feature sequences using modality-specific bidirectional GRUs, multimodal fusion, and prototype-based classification with an optional short user adaptation. Microphones achieve the strongest single-sensor performance, with median macro $F_1$ scores of 97.2\% under leave-one-food-out (LOFO) and 95.7\% under leave-one-subject-out (LOSO) evaluation after user adaptation. Multimodal fusion reaches 98.0\% under LOFO and 97.3\% under adapted LOSO. We demonstrate CHOMP's performance under three acoustic-interference conditions and within a cafeteria setting. Our results establish earphones as a practical platform for everyday CSP monitoring and jaw-function assessment.

\end{abstract}

\begin{teaserfigure}
  \includegraphics[width=\textwidth]{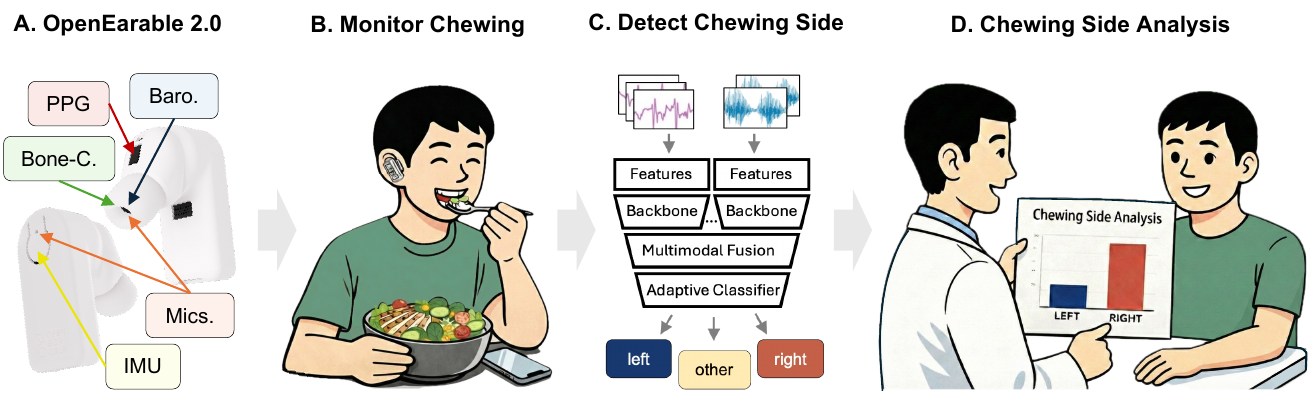}
  \centering
  \caption{CHOMP pipeline: A. Two OpenEarable~2.0 devices collect data from five sensing modalities. B. A user wears the devices during a meal while chewing activity is recorded. C. Feature sequences from the different sensing modalities are processed by modality-specific backbones, fused into a shared representation, and classified by a user-adapted classifier as left chewing, right chewing, or other activity. D. An orthodontist discusses the resulting chewing-side analysis with the user.}
  \Description{Overview of the CHOMP pipeline. A. Two OpenEarable~2.0 devices provide bilateral sensing using inner and outer microphones, a bone-conduction microphone, an IMU, PPG, and a pressure sensor. B. A user wears the devices during a meal while chewing activity is recorded. C. Feature sequences from the sensing modalities are processed by modality-specific backbones, fused into a shared representation, and classified by a user-adapted classifier as left chewing, right chewing, or other activity. D. An orthodontist discusses the resulting chewing-side analysis with the user using a chart comparing chewing on the left and right sides.}
  \label{fig:teaser}
\end{teaserfigure}

\maketitle

\section{Introduction}\label{sec: Intro}

Chewing-side preference (CSP) describes the habitual tendency to chew predominantly on one side of the jaw~\cite{santana-mora_temporomandibular_2013, jiang_analysis_2015}. It has been shown to be both a risk factor~\cite{barcellos_prevalence_2011, tiwari_chewing_2017} for developing temporomandibular disorders~(TMDs) and a behavioral manifestation of existing TMDs~\cite{lopez-cedrun_jaw_2017, yap_sleeping_2024}. TMDs affect approximately one-third of the global population~\cite{alqutaibi_global_2025, zielinski_meta-analysis_2024}. Despite their prevalence, TMDs are primarily diagnosed and monitored through self-reports and clinical examinations~\cite{national_academies_of_sciences_prevalence_2020, diernberger_self-reported_2008}, providing only episodic insight into everyday jaw function and its development over time. Objective, longitudinal measurements of chewing behavior could complement these assessments by revealing persistent asymmetric chewing patterns in daily life.
% Accordingly, at the beginning of the last decade, annual treatment and management costs were already estimated at USD 4 billion in the US alone~\cite{nih_facial_nodate}. 

Several wearable approaches have been proposed for chewing-side detection, using surface EMG electrodes~\cite{yamasakiObjectiveAssessmentActual2015}, microphones positioned around the face~\cite{nakamura_automatic_2020a,nakamura_automatic_2020b,nakamura_automatic_2021}, a headband~\cite{wang_wearable_2021}, or glasses~\cite{chung_glasses-type_2017}. However, most rely on dedicated sensing form factors not routinely worn in everyday life. In contrast, earphones are widely adopted and socially accepted, with true wireless stereo (TWS) devices now representing the dominant consumer form factor~\cite{hu_survey_2025}. Positioned close to the temporomandibular joint, they can capture complementary chewing-related signals, including jaw-related motion via inertial sensing~\cite{dai_poster_2025}, chewing acoustics via microphones~\cite{ketmalasiri_imchew_2024}, and subtle ear-canal deformations reflected in pressure changes~\cite{ando_canalsense_2017}, making them a promising platform for longitudinal, everyday chewing-side monitoring.

To this end, we present CHOMP (Chewing-Side Observation via Multimodal EarPhones), the first system for chewing-side detection using earphones. We employ OpenEarable~2.0~\cite{roddiger_openearable_2025}, a TWS earable research platform integrating inner and outer air-conduction microphones, a bone-conduction microphone, an inertial measurement unit (IMU), photoplethysmography~(PPG), and a pressure sensor. While these modalities have previously been shown to be informative for chewing detection~\cite{hu_survey_2025, roddiger_sensing_2022}, their individual suitability and complementary value for chewing-side detection remain largely unexplored. We therefore systematically characterize their contributions through multimodal fusion.

CHOMP processes paired-ear temporal feature sequences using modality-specific bidirectional GRUs with attentive temporal aggregation, followed by multimodal fusion and prototype-based classification. The prototype-based formulation additionally enables optional, lightweight user adaptation from a short calibration recording, following the gradient-free Bayesian prototype adaptation approach introduced in BayaHAR~\cite{burzer2026uncertainty}. We evaluate CHOMP in a study with 20 participants performing non-chewing activities and left- and right-sided chewing across ten mechanically diverse food items, a complete meal, and several acoustic-interference conditions. Using leave-one-food-out (LOFO) and leave-one-subject-out (LOSO) cross-validation, we assess generalization to unseen foods and participants. Our results show that microphones provide the strongest standalone performance, while multimodal fusion yields further improvement and short user-specific calibration enhances performance for unseen users. CHOMP also maintains high performance across three acoustic-interference conditions. Aggregating held-out predictions enables meal-level estimation of chewing-side asymmetry, while participants reported high perceived accuracy during a live cafeteria deployment.

Taken together, our key contributions are as follows: 

\begin{enumerate}
    \item We introduce CHOMP, the first system for chewing-side detection using earphones;
    \item We construct a mechanically diverse and balanced food set grounded in established food-science methods;
    \item We develop a multimodal paired-ear model architecture with modality-specific temporal modeling and prototype-based classification, enabling lightweight user adaptation from a short calibration recording;
    \item We systematically characterize five sensing modalities in a 20-participant study, showing that microphones perform strongest individually and multimodal fusion reaches median macro $F_1$ scores of up to $98.0\%$ in LOFO and $97.3\%$ in LOSO evaluation; and
    % \item We systematically characterize the suitability of five sensing modalities for earphone-based chewing-side detection through a study with 20 participants, showing that microphones provide the strongest standalone performance (LOFO: median $F_1 = 96.5\%$; LOSO: median $F_1 = 95.0\%$), while fusion with the bone-conduction microphone and the IMU yields complementary improvements (LOFO: median $F_1 = 97.2\%$; LOSO: median $F_1 = 96.9\%$); and
    \item We evaluate CHOMP beyond isolated-food classification through acoustic-interference experiments, meal-level chewing-side asymmetry estimation, and a live cafeteria deployment.
    % \item We demonstrate the robustness of our approach through evaluations across diverse foods, full meals, and environmental noise.
\end{enumerate}

% we introduce CHOMP, the first system for chewing-side detection using earphones; (2) we bu (3) we systematically characterize the suitability of five sensing modalities for earphone-based chewing-side detection through a study with 20 participants, showing that microphones provide the strongest standalone performance (LOFO: median $F_1 = 96.5\%$; LOSO: median $F_1 = 95.0\%$), while fusion with the bone-conduction microphone and the IMU yields complementary improvements (LOFO: median $F_1 = 97.2\%$; LOSO: median $F_1 = 96.9\%$); and (4) we demonstrate the robustness of our approach through evaluations across diverse foods, full meals, and environmental noise.

% (2) we collected data from 20 participants across five sensing modalities (microphones, bone-conduction microphone, IMU, PPG, pressure sensor) and show that microphones provide the strongest single-sensor unit performance (LOFO: median $F_1 = 94.5\%$, LOSO: median $F_1 = 92.6\%$) and that fusion with bone-conduction microphone and IMU further elevates performance (LOFO: median $F_1 = 97.7\%$, LOSO: median $F_1 = 95.4\%$)
\section{Related Work}\label{sec: Related Work}

% In the following, we first introduce the phenomenon of chewing-side preference (CSP) and its relation to temporomandibular disorders (TMDs) in \autoref{sec: rw_csp_tmd} and then review prior wearable-based approaches for detecting the chewing-side and TMDs in \autoref{sec: rw_detecting_csp_and_tmd}. Furthermore, we show how previous efforts on chewing detection with earables inform this work in \autoref{sec: rw_chewing_with_earables}.

We first review the clinical relevance of chewing-side preference (CSP) and limitations of current temporomandibular disorder (TMD) assessment. We then discuss prior wearable approaches to chewing-side detection and broader work on chewing detection with earables that informs the design of CHOMP.

\subsection{Chewing-Side Preference and Temporomandibular Disorders}\label{sec: rw_csp_tmd}

Chewing-side preference is highly prevalent across populations~\cite{barcellos_prevalence_2011, martinez-gomis_relationship_2009, rovira-lastra_peripheral_2016} and can already be observed in children~\cite{mc_donnell_chewing_2004}, yet individuals are often unaware of their own chewing-side preference~\cite{tiwari_chewing_2017}. Beyond contributing to asymmetric tooth wear \cite{barcellos_prevalence_2011}, long-term CSP has been associated with morphological alterations in craniofacial structures \cite{jiang_assessment_2015}, nasal airway asymmetry \cite{pecci-lloret_observational_2025}, and an increased likelihood of hearing loss \cite{lee_unilateral_2019}.

CSP is both a risk factor~\cite{barcellos_prevalence_2011, tiwari_chewing_2017} for developing temporomandibular disorders (TMDs) and a behavioral manifestation in patients with existing TMDs~\cite{lopez-cedrun_jaw_2017, yap_sleeping_2024}. TMDs comprise more than thirty specific conditions~\cite{national_academies_of_sciences_prevalence_2020} and encompass diverse physical symptoms, including joint pain, clicking, and limited motion~\cite{maini_temporomandibular_2025}, but also psychological correlates such as depression, anxiety, and sleep disturbances~\cite{park_association_2025}. Although symptoms resolve spontaneously in up to 40\% of patients~\cite{scrivani_temporomandibular_2008}, high chronic pain is nearly four times as common among individuals with TMDs~\cite{national_academies_of_sciences_prevalence_2020}. The relevance of TMDs is further amplified by the frequent use of the temporomandibular joints (TMJs), which perform approximately 2,000 opening and closing movements each day~\cite{national_academies_of_sciences_prevalence_2020}.
% Etiological factors are multifacetted, involving genetics, trauma, stress, and socioeconomic conditions \cite{minervini_economic_2023}. 
% TMDs are present in around a third of the global population \cite{alqutaibi_global_2025, zielinski_meta-analysis_2024}, with a pronounced gender imbalance toward women \cite{bueno_gender_2018}. At the beginning of the last decade, annual treatment and management costs were already estimated at approximately USD 4 billion in the US alone \cite{nih_facial_nodate}. 

The \textit{Diagnostic Criteria for Temporomandibular Disorders (DC/TMD)} are considered the current gold standard for research and clinical assessment~\cite{schiffman_diagnostic_2014}. In practice, however, they are not applied consistently, with practitioners reporting difficulties in accurate diagnosis~\cite{national_academies_of_sciences_prevalence_2020}, while their application has been considered impractical in routine clinical settings~\cite{steenks_temporomandibular_2018}. Current assessment relies predominantly on self-reports and clinical examinations~\cite{national_academies_of_sciences_prevalence_2020, diernberger_self-reported_2008}, although imaging techniques are starting to play a role \cite{talmaceanu_imaging_2018}. Consequently, longitudinal monitoring of everyday jaw function is rarely performed, prompting calls for the development of innovative approaches~\cite{national_academies_of_sciences_prevalence_2020}.

A wearable solution for monitoring CSP could complement current assessment by enabling longitudinal, quantitative measurements of chewing behavior in everyday life. Such measurements could provide behavioral insights that are difficult to obtain through episodic clinical examinations. Persistent or emerging chewing asymmetries could thereby help characterize potentially maladaptive patterns and track changes in chewing behavior over time, supporting more informed and personalized management.

\subsection{Detecting CSP and TMD using Wearable Approaches}\label{sec: rw_detecting_csp_and_tmd}

Motivated by the limitations of conventional clinical assessment, several wearable systems have explored objective measurements of chewing behavior in everyday settings. For direct TMD assessment, \citet{tsuji_heartmd_2025} employed earphones to measure maximum mouth opening and discriminate TMD patients from healthy controls with an AUC of 70.0\%. All other wearable approaches focused on identifying CSP, often with regard to TMD~\cite{wang_wearable_2021, yamasakiObjectiveAssessmentActual2015, kim_earchew_2022} and are comprehensively presented in \autoref{tab: csd_studies}. \citet{yamasakiObjectiveAssessmentActual2015} presented the first wearable approach using surface EMG electrodes placed over the left and right masseter muscles. They differentiated left- and right-sided chewing for two of three tested foods based on contraction amplitudes. A series of microphone-based approaches by ~\citet{nakamura_automatic_2020a,nakamura_automatic_2020b,nakamura_automatic_2021} employed Mel-Frequency Cepstral Coefficients (MFCCs) and cross-correlation features. Using arrays of up to five microphones positioned around the neck, throat, and mouth, they achieved $F_1$ scores of up to 81.0\% for classifying left, right, and front chewing and swallowing. Conceptually related, \citet{kim_earchew_2022} describe preliminary work on chewing-side detection using an in-ear microphone. The evaluation was restricted to a single participant chewing almonds, with an achieved accuracy of 96.0\%. Motion-based approaches leveraged chewing-related movements of the temporalis muscle. \mbox{\citet{wang_wearable_2021}} used a headband with bilateral IMUs and employed personalized models, thereby achieving 84.8\% accuracy in five-fold CV and 81.4\% in a leave-one-food-out CV. An approach employing a glasses-type wearable has been presented by \citet{chung_glasses-type_2017}, who used load cells embedded in the hinges to measure temporalis muscle activity through amplified hinge forces to discriminate left and right chewing within six facial activities in total. Using an SVM with statistical, temporal, and spectral features, the authors achieved $F_1$ scores of 89.5\% for left and right chewing in a leave-one-participant-out evaluation.

\begin{table*}[t]
    \centering
    \footnotesize
    \caption{Overview of experimental configurations in previous chewing-side detection studies in comparison with this work.}
    \begin{tabular}{p{2.3cm}p{2.5cm}p{4.2cm}>{\centering\arraybackslash}p{1.2cm}>{\centering\arraybackslash}p{0.7cm}>{\centering\arraybackslash}p{0.7cm}>{\centering\arraybackslash}p{1.8cm}}
    \toprule
    \textbf{Study} & \textbf{Embodiment} & \textbf{Sensor Units Employed} & \textbf{Wireless} &\textbf{Foods} & \textbf{Meal} & \textbf{Audio Robust.} \\
    \midrule
    \citet{yamasakiObjectiveAssessmentActual2015} & Surface Electrodes & EMG & No & 3 & No & No \\\addlinespace[2pt] % Proper spacing between 
    \citet{nakamura_automatic_2020a} & Neck Clip & Microphones & No & 3 & No & No \\
    \addlinespace[2pt] % Proper spacing between 
    \citet{nakamura_automatic_2020b} & Neck Clip & Microphones & No & 5 & No & No \\
    \addlinespace[2pt] % Proper spacing between 
    \citet{nakamura_automatic_2021} & Neck Clip, Head Clips & Laryngophones, Microphone & No & 5 & No & No \\
    \addlinespace[2pt] % Proper spacing between 
    \citet{kim_earchew_2022} & In-Ear Insertion & Microphone & \textit{N/A} & 1 & No & No \\
    \addlinespace[2pt] % Proper spacing between 
    \citet{wang_wearable_2021} & Headband & IMUs & Yes & 8 & No & No \\
    \addlinespace[2pt] % Proper spacing between 
    \citet{chung_glasses-type_2017} & Glasses & Load Cells on Hinges & No & 3 & No & No \\
    \addlinespace[2pt] % Proper spacing between 
    \midrule
    CHOMP & Earphones & Microphones, Bone-Conduction Microphone, IMU, PPG, Pressure Sensor & Yes & 11 & Yes & Yes \\
    \addlinespace[2pt] % Proper spacing between 
    \bottomrule
    \end{tabular}
    \label{tab: csd_studies}
\end{table*}

Despite these efforts, existing systems face several limitations. First, most approaches rely on dedicated sensing form factors or sensor placements that are not routinely worn in everyday life. Second, with the exception of \citet{wang_wearable_2021}, previous studies evaluate no more than five foods, and none examines chewing-side detection with heterogeneous meals or under acoustic interference. Third, most prior bilateral systems rely on tethered or jointly mounted sensors, enabling tight temporal alignment between the two sides and the use of sample-level cross-correlational features. In independent TWS earphones, however, the inter-ear propagation-time differences that could encode directional information are small, such that even minor synchronization offsets between devices can substantially distort delay estimates \cite{hunn_introducing_2024}. Consequently, such features are difficult to exploit reliably without dedicated inter-ear synchronization.

\subsection{Chewing Detection with Earables}\label{sec: rw_chewing_with_earables}

A substantial body of research has investigated chewing detection more broadly using earable devices~\cite{roddiger_sensing_2022, bell_automatic_2020, vu_wearable_2017, he_comprehensive_2020}. Most studies focus on detecting chewing episodes and distinguishing them from non-chewing activities~\cite{bedriEarBitUsingWearable2017, ketmalasiri_imchew_2024, lotfi_comparison_2020}, while others more specifically address chew counting ~\cite{ketmalasiri_imchew_2024}, snacking behavior ~\cite{bin_morshed_personalized_2022}, food-intake estimation~\cite{chugh_bitesense_2025}, or food-type recognition~\cite{amft_body_2009}. Across this literature, approaches primarily rely on microphones~\cite{nishimura_eating_2008, pasler_food_2011} or IMUs~\cite{ketmalasiri_imchew_2024, lotfi_comparison_2020}. Other sensing strategies include bone-conduction microphones~\cite{kondo_optimized_2019, shuzo_wearable_2010, bi_auracle_2018}, proximity sensing~\cite{bedriEarBitUsingWearable2017, ID86_bedri_detecting_2015, taniguchi_earable_2018}, PPG~\cite{papapanagiotou_novel_2017}, and pressure sensing~\cite{hossain_ear_2023}. Combining sensing modalities has further been shown to improve chewing-detection performance~\cite{lotfi_comparison_2020, papapanagiotou_novel_2017}. Since these approaches often rely on technically demanding configurations or non-standard sensor placements, many systems are built on custom hardware. 
% Commercial earables such as eSense \cite{kawsar_earables_2018} appear only occasionally \cite{ketmalasiri_imchew_2024, lotfi_comparison_2020, amft_body_2009, chugh_bitesense_2025, bin_morshed_personalized_2022, gao_ihear_2016}, thus limiting the transferability of existing findings to everyday consumer devices. 
This heterogeneity in setups and device configurations has produced an equally diverse range of feature representations. Prior work employs statistical (e.g., mean, variance, energy \cite{lotfi_comparison_2020, bi_auracle_2018}), temporal (e.g., zero-crossing rate \cite{hossain_ear_2023}), kinematic (e.g., jerk \cite{chugh_bitesense_2025}), and spectral or cepstral representations, particularly MFCCs~\cite{nishimura_eating_2008, kondo_optimized_2019, lotfi_comparison_2020}. 

Taken together, prior work informs CHOMP in two ways. First, evidence across multiple sensing mechanisms supports a systematic evaluation of the complementary modalities available in OpenEarable~2.0. Second, the diversity of feature representations reflects the different physical phenomena captured by each sensor and motivates modality-specific features, while the synchronization constraints of independent TWS devices require bilateral fusion that does not depend on precisely synchronized sample-level correlations.

\begin{figure}[t]
    \centering
    \includegraphics[width=1\linewidth]{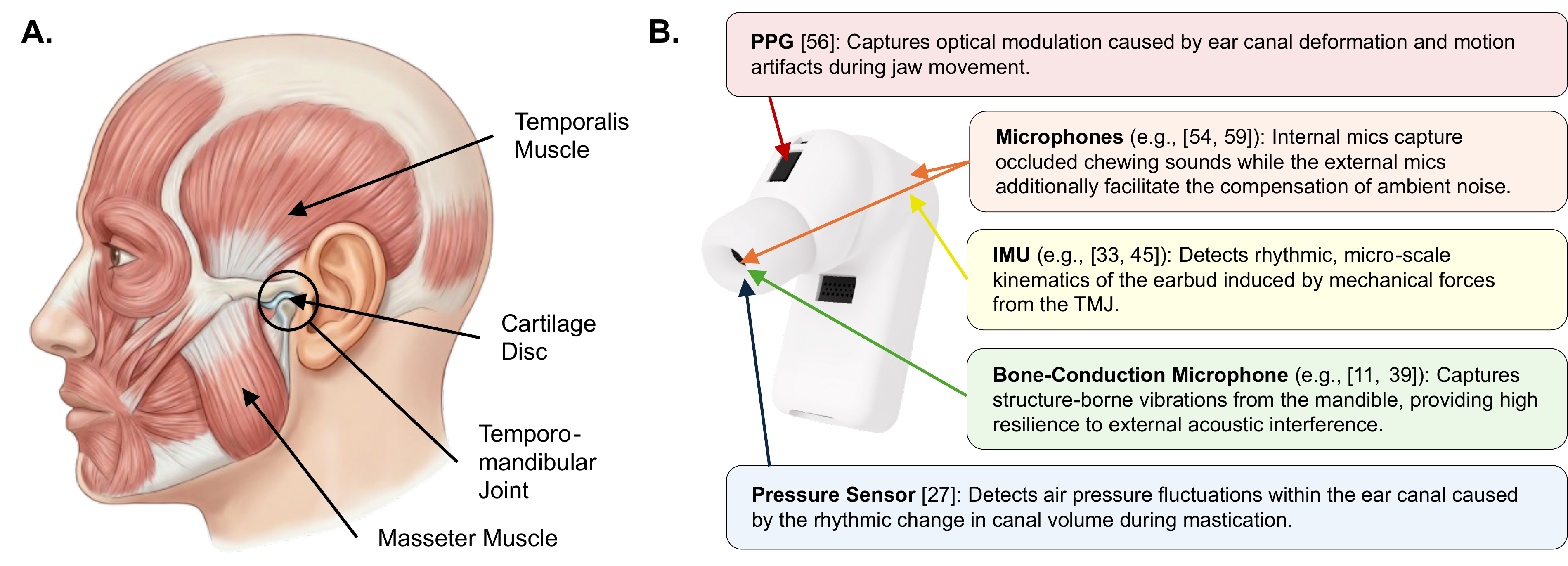} % Change to your image filename
    \caption{Anatomical and sensing basis of CHOMP. A. Anatomical structures relevant to TMD are located in close proximity to the ear (adapted from \cite{university_of_leeds_improving_2025}); B. Sensing modalities integrated in OpenEarable 2.0 and their rationale for effective chewing-side detection, along with supporting literature on earphone-based chewing recognition.}
    \Description{A. The anatomical structures relevant to TMD are located in close proximity to the ear (adapted from \cite{university_of_leeds_improving_2025}). Highlighted are the Temporalis Muscle, the Temporomandibular Joint and its Cartilage Disc, and the Masseter Muscle; B. Sensing modalities integrated in OpenEarable 2.0 and their rationale for effective chewing-side detection, along with supporting literature on earphone-based chewing recognition. 
    
    PPG \cite{papapanagiotou_novel_2017}: Captures optical modulation caused by ear canal deformation and motion artifacts during jaw movement.; 
    Microphones (e.g., \cite{nishimura_eating_2008,pasler_food_2011}): Internal mics capture occluded chewing sounds while the external mics additionally facilitate the compensation of ambient noise
    IMU (e.g., \cite{ketmalasiri_imchew_2024, lotfi_comparison_2020}): Detects rhythmic, micro-scale kinematics of the earbud induced by mechanical forces from the TMJ
    Bone-Conduction Microphone (e.g., \cite{kondo_optimized_2019, bi_auracle_2018}): Captures structure-borne vibrations from the mandible, providing high resilience to external acoustic interference.
    Pressure Sensor \cite{hossain_ear_2023}: Detects air pressure fluctuations within the ear canal caused by the rhythmic change in canal volume during mastication}
    \label{fig: Anatomical_Rationale}
\end{figure}

\section{Methodology}\label{sec: Methodology}

% In the following, we outline the experimental methodology. First, we describe the employed measurement device, OpenEarable 2.0 \cite{roddiger_openearable_2025}, and its integrated sensors in \autoref{sec: openearable}. Subsequently, we present the experimental protocol used for the data collection in \autoref{sec:experimental_protocol}, including a detailed account of each experimental block and the corresponding design rationale.

We outline the sensing platform and experimental protocol underlying the collection of the CHOMP dataset.

\subsection{OpenEarable 2.0}\label{sec: openearable}

Participants wore two OpenEarable~2.0 devices, one on each ear, throughout all recordings. We used their multimodal sensing capabilities to capture complementary chewing-related signals. Specifically, we recorded the 3-axis accelerometer and 3-axis gyroscope of the IMU (100~Hz), the pressure sensor (100~Hz), the photoplethysmography (PPG) sensor (50~Hz), the 3-axis bone-conduction microphone (1.6~kHz), and the inner and outer air-conduction microphones (48~kHz). To maintain stable placement, the devices were secured with wingtips. \autoref{fig: Anatomical_Rationale} illustrates OpenEarable~2.0 and its sensing modalities.

% \begin{figure*}[t]
%     \centering
%     \includegraphics[width=\linewidth]{figs/OpenEarable_PLACEHOLDER.png} % Change to your image filename
%     \caption{OpenEarable: PLACEHOLDER - do we even include sth alike?}
%     \Description{TODO}
%     \label{fig: OpenEarable}
% \end{figure*}

\subsection{Experimental Protocol}\label{sec:experimental_protocol}

\begin{figure*}[t]
    \centering
    \includegraphics[width=\linewidth]{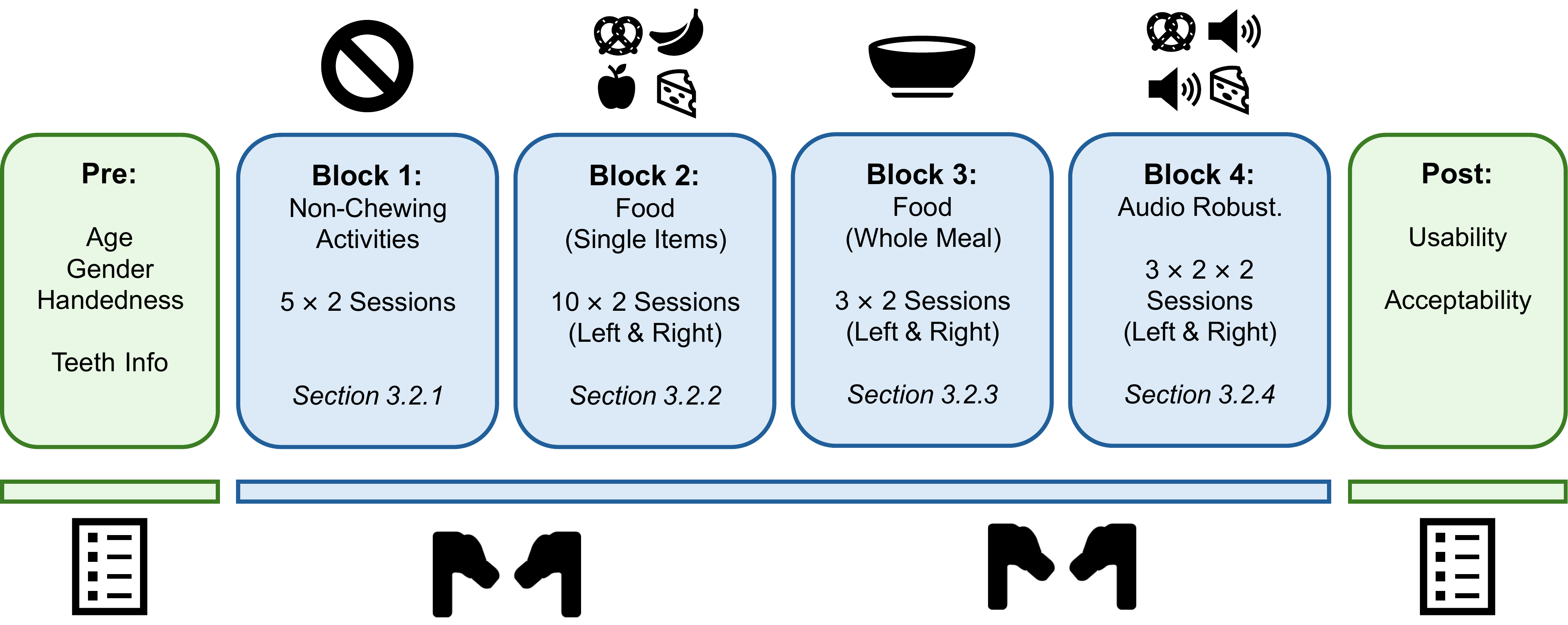} % Change to your image filename
    \caption{Experimental overview. Participants completed a pre-questionnaire (see \autoref{app: pre_quest}), performed four experimental blocks while wearing OpenEarable 2.0 devices, and then filled out a post-questionnaire (see \autoref{app: post_quest}). Block 1 included five non-chewing activities (see \autoref{sec: selection_non-chew}) while Blocks 2–4 comprised different chewing tasks on the left and right side: consumption of ten single food items (see \autoref{sec: selection_foods}), a whole meal (see \autoref{sec: whole_meal}), and two selected foods under three acoustic-interference conditions (see \autoref{sec: external_validity}).}
    \Description{Experimental overview. Participants completed a pre-questionnaire (see \autoref{app: pre_quest}), performed four experimental blocks while wearing OpenEarable 2.0 devices, and then filled out a post-questionnaire (see \autoref{app: post_quest}). Block 1 included five non-chewing activities (see \autoref{sec: selection_non-chew}) while Blocks 2–4 comprised different chewing tasks on the left and right side: consumption of ten single food items (see \autoref{sec: selection_foods}), a whole meal (see \autoref{sec: whole_meal}), and two selected foods under three audio noise conditions (see \autoref{sec: external_validity}).
    
    In the figure, a summary scheme of the experiment is shown that reads as follows:
    
    Pre: Age, Gender, Handedness, Teeth Info, via Questionnaire;
    Block 1: Non-Chewing Activities, 5 x 2 Sessions, via OpenEarable 2.0 (see \autoref{sec: selection_non-chew});
    Block 2: Food (Single Items), 10 x 2 Sessions, Left and Right, via OpenEarable 2.0 (see \autoref{sec: selection_foods});
    Block 3: Food (Whole Meal), 3 x 2 Sessions, Left and Right, via OpenEarable 2.0 (see \autoref{sec: whole_meal});
    Block 4: Audio Robustness, 3 x 2 x 2 Sessions, Left and Right, via OpenEarable 2.0 (see \autoref{sec: external_validity});
    Post: Usability, Acceptability, via Questionnaire;}
    \label{fig: Experimental Scheme}
\end{figure*}

The experimental protocol consisted of four blocks, preceded by a pre-questionnaire and followed by a post-questionnaire. A schematic overview is provided in \autoref{fig: Experimental Scheme}. The pre-questionnaire collected demographic information (age, gender, handedness) as well as relevant dental details (see \autoref{app: pre_quest}). Following data collection, the post-questionnaire assessed the usability and acceptability of OpenEarable~2.0 for chewing-side detection (see \autoref{app: post_quest}).

Each session consisted of a 30-second recording, with multiple sessions conducted within each experimental block. For chewing sessions, participants were instructed to chew on the designated left or right side throughout the recording. Sessions within each block were randomized to minimize order effects. The study protocol was approved by the university's ethics committee. The design rationale and structure of each block are detailed below.

\subsubsection{Non-Chewing Activities}\label{sec: selection_non-chew}

To represent the non-chewing class and challenge the model with potentially confounding activities, we selected five conditions informed by prior chewing-detection studies. These included \textit{sitting still}~\cite{lotfi_comparison_2020, ketmalasiri_imchew_2024, farooq_comparative_2015, farooq_novel_2016} and \textit{drinking and swallowing}~\cite{lotfi_comparison_2020, ketmalasiri_imchew_2024, liu_intelligent_2012, gao_ihear_2016}, although the latter did not necessarily occupy the entire 30-second recording. From \citet{ketmalasiri_imchew_2024}, we adapted \textit{head movements} from side-to-side motion and generalized emotional facial expressions into \textit{drawing faces}. \textit{Reading aloud} reflected prior use of reading and speaking tasks~\cite{lotfi_comparison_2020, ketmalasiri_imchew_2024, liu_intelligent_2012, gao_ihear_2016, farooq_comparative_2015, farooq_novel_2016}. We did not include \textit{walking}~\cite{farooq_comparative_2015, farooq_novel_2016}, as the study focused exclusively on seated eating (see \autoref{sec: limitations}), nor \textit{watching a movie}, as potential acoustic interference from media consumption was evaluated separately in the audio-robustness block (see \autoref{sec: external_validity}). Each activity was recorded in two sessions, yielding ten non-chewing sessions per participant.

% \begin{figure}[t]
%     \centering
%     \includegraphics[width=0.5\linewidth]{figs/Presser_almond_cheese.pdf}
%     \caption{Texture profile analysis of semi-hard cheese (left) and almonds (right) using the double-compression test with a 20 mm diameter cylinder probe.}
%     \Description{TO EXCHANGE ANYWAYS}
%     \label{fig: presser}
% \end{figure}

\subsubsection{Food (Single Items)}\label{sec: selection_foods}

\begin{figure*}[t]
    \centering

    \begin{subfigure}[t]{0.42\textwidth}
        \centering
        \caption{Texture Profile Analysis Setup.}
        \label{fig: presser}
        \includegraphics[width=\linewidth]{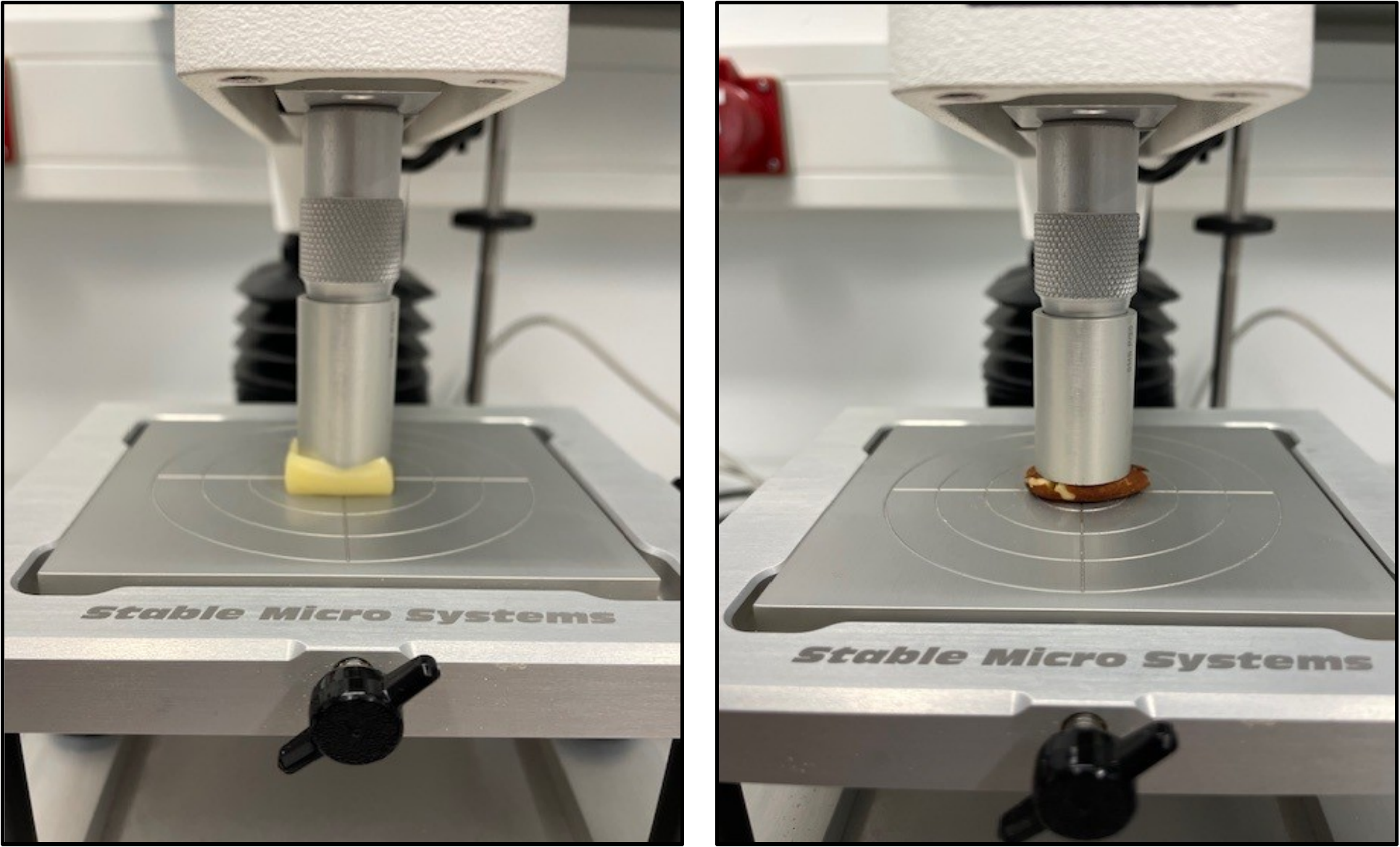}
    \end{subfigure}
    \hfill
    \begin{subfigure}[t]{0.54\textwidth}
        \centering
        \caption{Measured Food Firmness.}
        \label{fig: firmness}
        \includegraphics[width=\linewidth]{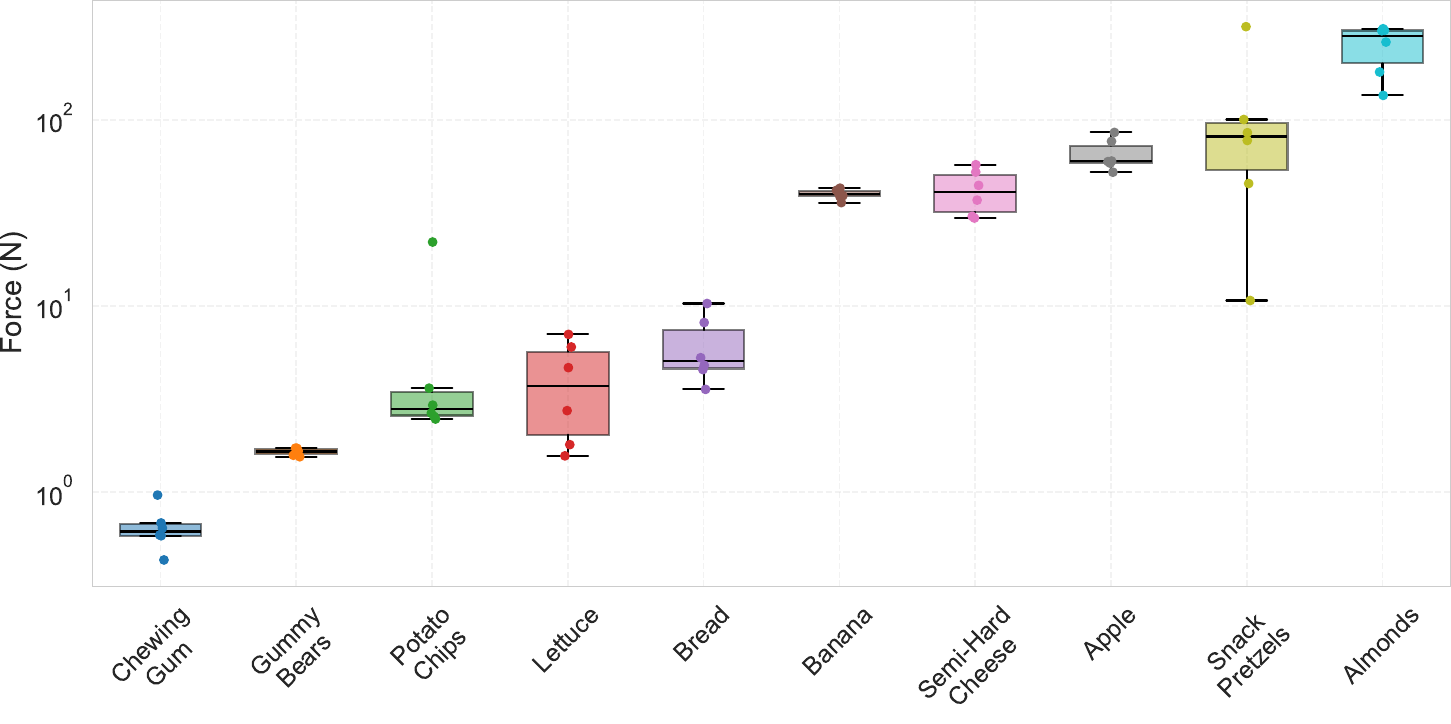}
    \end{subfigure}

    \caption{Instrumental characterization of the selected food items. A. Texture Profile Analysis of semi-hard cheese and almonds using the double-compression test with a 20\,mm diameter cylinder probe. B. Firmness of the ten food samples ($N = 6$ each).}
    \Description{The figure shows the instrumental characterization of the selected foods. Panel A contains two photographs of the texture analyzer compressing semi-hard cheese and an almond with a cylindrical probe. Panel B shows boxplots of measured firmness for the ten food items on a logarithmic force scale, spanning from soft foods such as chewing gum to substantially firmer foods such as snack pretzels and almonds.}
    \label{fig: food_texture_analysis}
\end{figure*}

\begin{figure}[b]
    \centering
    \includegraphics[width=1\linewidth]{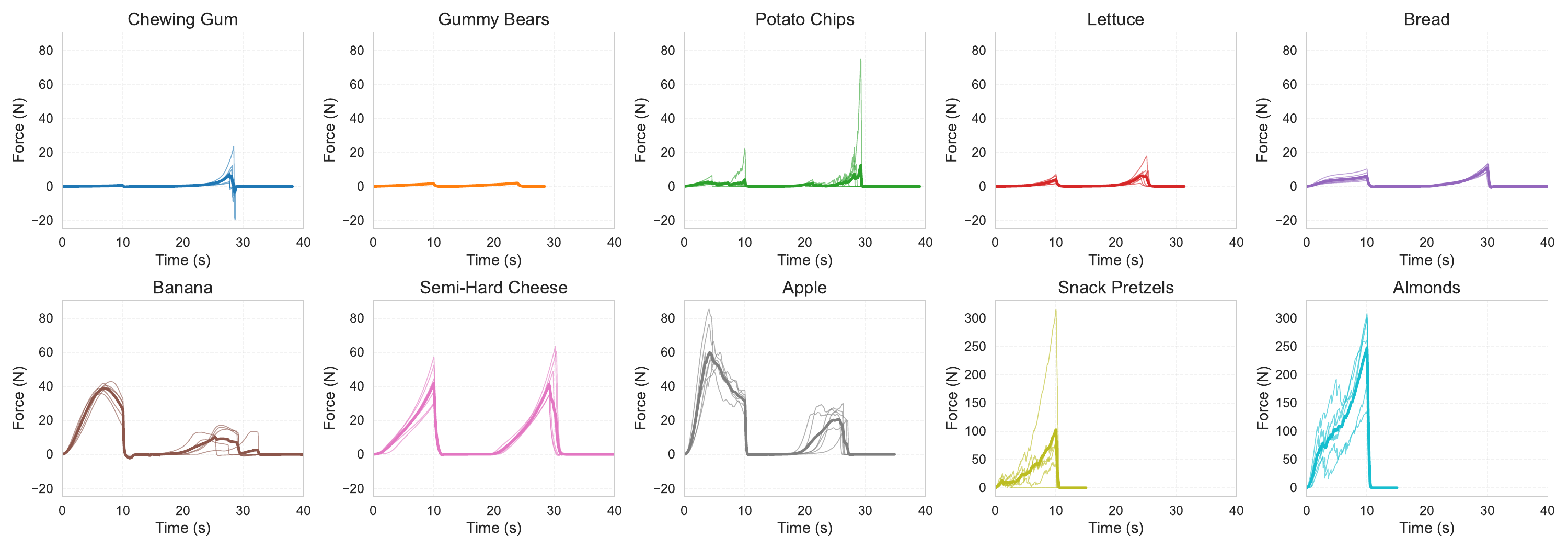}
    \caption{Force–time curves from the Texture Profile Analysis of the ten selected food samples ($N=6$ each), ordered by mean firmness. The y-axis scale differs for snack pretzels and almonds because of their substantially higher peak forces. }
    % \caption{Force-time curves recorded for the ten different food samples ($N = 6$ each), sorted by mean force applied. Note that the y-axis had to be scaled differently for the two hardest foods.}
    \Description{The figure shows force over time for six samples of each of the ten selected foods. Each panel corresponds to one food, with the individual force-time curves overlaid. Softer foods exhibit lower peak forces, while harder foods such as snack pretzels and almonds produce substantially larger peaks. The y-axis range is increased for snack pretzels and almonds to accommodate their higher forces.}
    \label{fig: curves}
\end{figure}

Our choice of foods was guided by prior chewing studies and complemented by an instrumental characterization of their mechanical properties to ensure diversity across the selected set. Texture Profile Analysis (TPA)~\cite{shikamaTextureAnalysisFood} is a standard method based on a double-compression ("two-bite") test, in which a sample is compressed twice to emulate the mechanical action of chewing \cite{bourne1978tpa}. Each candidate food was tested using a 20\,mm cylindrical compression probe (Stable Micro Systems, Godalming, UK), shown in \autoref{fig: presser}, mounted on a texture analyzer (TA.XTplus, Stable Micro Systems) equipped with a 30\,kg load cell. Following established procedure~\cite{STEMLER_2022100442}, every sample was compressed twice to a fixed compression distance of 5\,mm at a test speed of 5\,mm/s (trigger force 0.049\,N), and the applied force was recorded over time. The resulting force-time curves are shown in \autoref{fig: curves}. For almonds and snack pretzels, fracture during the first compression prevented a meaningful second compression. Because several standard TPA measures require an intact second cycle, we selected \textit{firmness}, defined as the peak force during the first compression, as the texture metric that could be computed consistently across all foods.

\begin{figure*}[t]
    \centering
    \includegraphics[width=\linewidth]{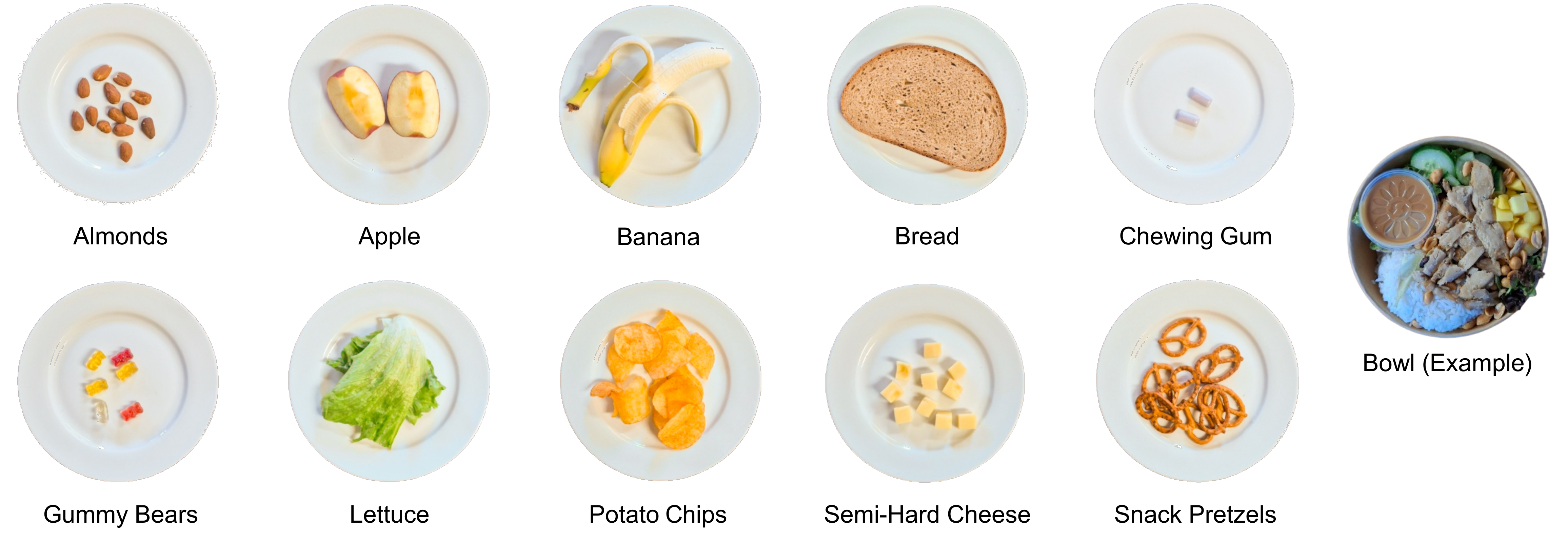} % Change to your image filename
    % \caption{Selection of different single food items (left) and an example of a whole meal (right).}
    \caption{Food items used in the experimental protocol. The ten single food items used in the isolated-food recordings are shown on the left, with an example of the customizable whole meal shown on the right.}
    \Description{Selection of different single food items (left) and an example of a whole meal (right). Almonds, Apple, Banana, Bread, Chewing Gum, Gummy Bears, Lettuce, Potato Chips, Semi-Hard Cheese, Snack Pretzels, Bowl (Example)}
    \label{fig: Food Selection}
\end{figure*}

The resulting firmness values shown in \autoref{fig: firmness} span a broad and balanced range, confirming the textural diversity of the food set and supporting our final selection of the following ten foods: \textit{almonds}~\cite{taniguchi_earable_2018, wang_wearable_2021}, \textit{snack pretzels}~\cite{lotfi_comparison_2020, ketmalasiri_imchew_2024, wang_wearable_2021}, \textit{potato chips}~\cite{lotfi_comparison_2020, chugh_bitesense_2025, bin_morshed_personalized_2022, ketmalasiri_imchew_2024, pasler_food_2011, pasler_food_2014, amft_body_2009, nishimura_eating_2008, gao_ihear_2016, khan_ihearken_2022}, \textit{apple}~\cite{ketmalasiri_imchew_2024, nakamura_automatic_2020b, pasler_food_2011, pasler_food_2014, lopez-meyer_detection_2010, amft_body_2009, gao_ihear_2016, sazonov_sensor_2012, khan_ihearken_2022}, \textit{gummy bears}~\cite{wang_wearable_2021}, \textit{chewing gum}~\cite{nakamura_automatic_2020a}, \textit{banana}~\cite{lotfi_comparison_2020, nishimura_eating_2008, khan_ihearken_2022}, \textit{lettuce}~\cite{lotfi_comparison_2020, amft_body_2009, nishimura_eating_2008}, \textit{bread}~\cite{ketmalasiri_imchew_2024, wang_wearable_2021, lopez-meyer_detection_2010, amft_body_2009, gao_ihear_2016, sazonov_sensor_2012}, and \textit{semi-hard cheese}~\cite{chugh_bitesense_2025}. To further diversify eating mechanics, five items were typically consumed by biting from a larger piece, whereas the remaining five could be consumed as individual portions. A photographic overview of all selected foods as used in the experiment is provided in \autoref{fig: Food Selection}. Each food was recorded once on the left and once on the right side, yielding 20 sessions per participant.

% \begin{figure}[t]
%     \centering
%     \includegraphics[width=0.5\linewidth]{figs/tpa_firmness_only_vertical.pdf}
%     \caption{Firmness of the ten different food samples ($N = 6$ each).}
%     \Description{TO EXCHANGE ANYWAYS}
%     \label{fig: firmness}
% \end{figure}

\subsubsection{Food (Whole Meal)}\label{sec: whole_meal}

To extend the evaluation beyond isolated food items, participants also consumed a heterogeneous meal. We selected a customizable bowl to accommodate individual food preferences while maintaining a common meal structure across participants. Each bowl comprised salad, rice, a crunchy element, a hard and a soft vegetable, and a protein, with the available options detailed in \autoref{app: bowl_characteristics}. Three recordings were collected for each chewing-side, yielding six sessions per participant.

\subsubsection{Acoustic Interference}\label{sec: external_validity}

To assess CHOMP under acoustic interference, we recorded additional eating sessions with two foods, semi-hard cheese and snack pretzels. Repeating all ten foods under each condition would have imposed substantial participant burden, so we selected these two foods to provide contrasting eating characteristics. We evaluated three acoustic-interference conditions. First, pop music was played through the earphones to reflect a typical use case. Second, dining hall ambience (mean sound level: 61.8\,dB) was played through an external speaker positioned in front of the participant to emulate a social eating environment\footnote{\href{https://www.youtube.com/watch?v=Xd7NyIhzAlg}{Dining Hall Ambiance Video}}. Third, construction noise (mean sound level: 71.3\,dB) was presented through the same setup as a more intense source of acoustic interference\footnote{\href{https://www.youtube.com/watch?v=AB4Ov9t4aq4}{Construction Noise Video}}. For each condition, both foods were consumed on each side, yielding 12 sessions per participant.

\subsubsection{Additional Procedures}\label{sec:further_measures}

To assess potential placement effects, participants removed and reattached the OpenEarable~2.0 devices between 10\% of sessions. These reattachments occurred at varying points throughout the experiment without a fixed schedule. The experimenter additionally annotated pauses for acquiring a new piece of food and swallowing events on a synchronized smartphone.

\section{Model Pipeline}
\label{sec:algorithm}

We introduce the CHOMP model pipeline by describing sensor data preprocessing in \autoref{sec: preprocessing} and feature extraction in \autoref{sec: feature_extraction}. Subsequently, we present the CHOMP model architecture and training procedure in \autoref{sec: model_architecture}. Finally, we describe the user adaptation in \autoref{sec: user-adaptation}.

\subsection{Preprocessing}
\label{sec: preprocessing}

\paragraph{Timestamps}
\label{sec:synchro}

As discussed in \autoref{sec: rw_detecting_csp_and_tmd}, independent TWS earphones do not inherently provide the precise inter-ear synchronization required for reliable sample-level delay estimation. Nevertheless, timestamps of the left and right earables must be aligned sufficiently to form paired windows. Since firmware version~2.2.2, OpenEarable 2.0 supports global timestamps, which removes the need for manual alignment. At the time of this study, firmware version~2.1.5 was the most recent release and did not provide this feature. Manual synchronization was therefore required. Each earable initializes its clock at power-on, creating an initial offset. To estimate this offset, an alignment signal of four Dirac impulses spaced one second apart, followed by four seconds of white noise, was recorded on both earables. Generalized Cross-Correlation with Phase Transform (GCC-PHAT) \cite{1162830} was applied to the outer-microphone signals, and the resulting correlation was smoothed with a Gaussian filter. For each uninterrupted power-on block, the temporal offset, determined as the $\arg\max$ of the filtered correlation, was used to align the two clocks. Clock drift was negligible and required no correction. Following alignment, paired sessions were restricted to the time interval available from both earables, and all channels were resampled to their target sampling frequencies. The microphones were downsampled from 48~kHz to 8~kHz, while the bone-conduction channels remained at 1.6~kHz. For smartphone annotations (see \autoref{sec:further_measures}), both earables streamed optical temperature data at 8~Hz to the smartphone to generate paired timestamps.

\paragraph{Filtering}

To isolate chewing-related frequency components while attenuating environmental noise and motion artifacts, we applied fourth-order zero-phase Butterworth filters after resampling. The selected accelerometer, gyroscope, PPG, and barometric-pressure channels were filtered from $0.1$ to $6.0$~Hz. This range captures the mastication rhythm, typically below 3~Hz, while retaining its immediate harmonics and being consistent with \cite{chugh_bitesense_2025, ketmalasiri_imchew_2024, hossain_ear_2023}. The bone-conduction channels were high-pass filtered at $0.1$~Hz, with their upper range limited by the 800~Hz Nyquist frequency, to retain both jaw motion and higher-frequency structural vibration \cite{khan_ihearken_2022}. The inner and outer microphones were likewise high-pass filtered at $0.1$~Hz after resampling, with their upper range limited by the 4~kHz Nyquist frequency.

\paragraph{Windowing}

For chewing-side detection, the choice of window size must reflect the temporal and spectral characteristics of mastication. Dominant chewing frequencies have been reported between 0.94 and 2.17~Hz~\cite{po_time-frequency_2011}, implying that 1~s windows are usually enough to capture one chewing cycle. We therefore segmented the synchronized streams into 1~s windows with 50\% overlap, corresponding to a stride of 0.5~s, retaining a physiologically meaningful temporal context while supporting responsive feedback and remaining consistent with prior work \cite{bin_morshed_personalized_2022, hossain_ear_2023}. Around each manually annotated interruption, such as swallowing or acquiring another piece of food, we defined an exclusion interval extending 0.5~s before and after the event to reduce the risk of label ambiguity. Any 1~s window overlapping this interval was discarded. Each remaining window inherited its participant, session activity, and instructed chewing-side. Windows from designated chewing activities were labeled \textit{left} or \textit{right} according to the instructed side, whereas windows from the dedicated non-chewing sessions were assigned to the \textit{other} class. These sessions comprised drinking and swallowing, head and facial movements, reading, and sitting still (see \autoref{sec:experimental_protocol}).

\subsection{Feature Extraction}
\label{sec: feature_extraction}

After preprocessing, we construct a compact, ordered feature sequence for each ear and sensor modality, preserving temporal order within the 1~s window while accommodating the distinct physical characteristics and sampling rates of the acoustic, vibration, motion, and physiological signals. \autoref{tab:sequence_representations} summarizes the contribution of each physical channel, any sensor-level features, and the resulting sequence shape for each sensor unit. 

\begin{table*}[t]
    \centering
    \footnotesize
    \caption{Per-ear feature-sequence representations for a 1~s window. $C$ is the number of physical channels, $T$ the number of sequence steps, and $F$ the total number of features per step after concatenating all channel-wise and sensor-level features.}
    \begin{tabular*}{\textwidth}{@{\extracolsep{\fill}} l l l l l}
        \toprule
        \textbf{Sensor Unit} & \textbf{Channels $C$} & \textbf{Sampling Rate} & \textbf{Features $F$ per Time Step} & \textbf{Input Shape $(T,F)$} \\
        \midrule
        Microphones & $2$ (inner, outer) & 8,000~Hz & $2 \times 13$ MFCCs $=26$ & $(98,26)$ \\
        Bone-Conduction Microphone & $3$ ($x,y,z$) & 1,600~Hz & $3 \times 13$ MFCCs $=39$ & $(95,39)$ \\
        IMU & $6$ (3 acc., 3 gyro.) & 100~Hz & 6 raw $+$ 6 differences $+$ 2 RMS $=14$ & $(100,14)$ \\
        PPG & $3$ (red, IR, green) & 50~Hz & 3 raw $+$ 3 differences $=6$ & $(50,6)$ \\
        Pressure Sensor & $1$ & 100~Hz & 1 raw $+$ 1 difference  $=2$ & $(100,2)$ \\
        \bottomrule
    \end{tabular*}
    \label{tab:sequence_representations}
\end{table*}

\paragraph{Acoustic and Vibration Modalities}

For the inner and outer microphones, as well as the three-axis bone-conduction microphone, we compute 13 Mel-Frequency Cepstral Coefficients (MFCCs) \cite{davis1980comparison} from 40 Mel bands for each physical channel. MFCCs provide a compact representation of the short-time spectral envelope and have previously been applied to acoustic and bone-conducted chewing signals \cite{nakamura_automatic_2021,li2019chewing,xinyi2022automatic}. We use Hann-windowed frames with a 10~ms hop and a frame length of 25~ms for the microphones, consistent with commonly used settings for short-time MFCC extraction \cite{abbasi2026ldsc}. For the 1.6~kHz bone-conduction signal, we increase the frame length to 52.5~ms to provide sufficient frequency resolution for the 40-band Mel filterbank at the lower sampling rate. This representation retains short-time spectral information while substantially reducing the dimensionality of the original signals.

\paragraph{Motion and Physiological Modalities}

For the IMU, PPG, and pressure sensor, we augment the raw values at each time step with the first temporal difference of each physical channel. The raw values preserve the measured signal trajectories, while the temporal differences emphasize local changes associated with chewing. For the IMU, we additionally compute separate root-mean-square (RMS) magnitudes across the accelerometer and gyroscope axes, providing compact measures of overall motion magnitude that are less sensitive to its distribution across individual axes.

\subsection{Model Architecture}
\label{sec: model_architecture}

\begin{figure*}[b]
\centering
\includegraphics[width=\textwidth]{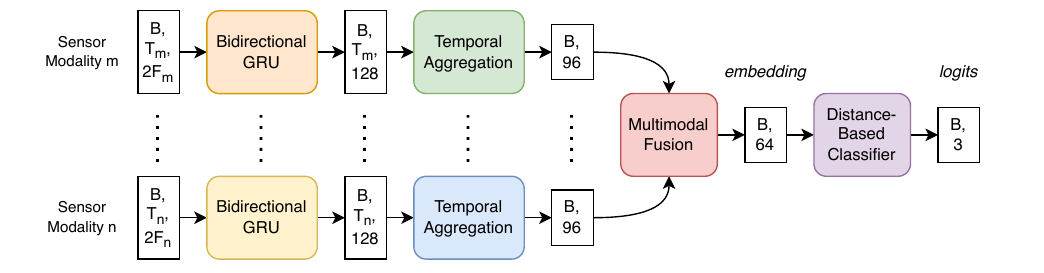}
\caption{CHOMP model architecture. For each sensor modality $m$, the paired left- and right-ear feature sequences $L_m$ and $R_m$ are concatenated along the feature dimension and processed by a modality-specific bidirectional GRU with temporal aggregation. Different colors indicate components with distinct learnable parameters. Each modality encoder produces a 96-dimensional representation, which is fused with the other modality representations into a shared 64-dimensional embedding. Classification is performed using squared Euclidean distances to three learned class prototypes for \textit{left}, \textit{right}, and \textit{other}, with a learned global temperature controlling the scale of the resulting logits.}
\Description{Architecture diagram of CHOMP showing multiple parallel sensor-modality branches. Each modality receives concatenated left- and right-ear feature sequences, which are processed by a modality-specific bidirectional GRU and temporal aggregation module to produce a 96-dimensional representation. The modality representations are combined by a multimodal fusion module into a 64-dimensional embedding. A distance-based classifier maps this embedding to three class logits. Different colors distinguish modules with separate learnable parameters. Tensor dimensions are shown between processing stages.}
\label{fig:arch}
\end{figure*}

For every modality $m$, the model receives synchronized left- and right-ear feature sequences $L_m,R_m\in\mathbb{R}^{T_m\times F_m}$, where $T_m$ and $F_m$ are given in \autoref{tab:sequence_representations}. We standardize each feature using statistics fitted exclusively on the training data, with separate statistics for the physical left and right ears. This accounts for persistent ear-specific offsets in the training data that could otherwise become unintended side cues. The same fixed statistics are subsequently applied to validation, test, and adaptation data to avoid data leakage. After standardization, the left- and right-ear sequences are concatenated along the feature dimension at each time step, yielding a sequence of shape $T_m\times 2F_m$. Each sensor modality is then processed by a separate encoder with the same architecture but distinct parameters, allowing the encoders to specialize to the temporal and spectral characteristics of their respective modalities. The resulting modality representations are fused into a shared embedding that forms the output of the backbone and the input to the three-class classifier for \textit{left}, \textit{right}, and \textit{other}.

\paragraph{Bidirectional GRU}

Chewing involves a short temporal progression of jaw motion, food fracture, and transmitted vibrations, requiring the model to capture how the bilateral relationship between the left- and right-ear features evolves throughout each window. We employ a GRU \cite{cho2014learning} rather than an LSTM \cite{hochreiter1997long}, as GRUs use fewer gates and parameters while providing sufficient modeling capacity for the relatively short sequences in our setting (see \autoref{tab:sequence_representations}). We use a bidirectional GRU so that each time step incorporates context from the complete 1~s window. Before recurrent processing, the concatenated $2F_m$-dimensional sequence is projected to 64 dimensions using a linear layer followed by a SiLU activation, providing all modality-specific GRUs with the same input dimensionality. A single-layer bidirectional GRU with 64 hidden units in each direction then produces $T_m$ hidden states of dimension 128.

\paragraph{Temporal Aggregation}

As the sensor modalities have different sequence lengths, their hidden-state sequences must be aggregated into fixed-dimensional representations before fusion. Temporally localized events, such as food fracture or pronounced jaw movement, may contain stronger side-related information than relatively stationary intervals. Since not all parts of a window are equally informative for chewing-side classification, we use attentive temporal aggregation rather than simple uniform averaging. A linear layer maps each hidden state to a scalar attention score, followed by a softmax over the temporal dimension to produce normalized weights that sum to one. Using these weights, we compute a 128-dimensional weighted mean and weighted standard deviation, capturing the central tendency and within-window variability of the hidden states. Their concatenation yields a 256-dimensional vector, which is projected to 96 dimensions using a linear layer, followed by layer normalization and a SiLU activation. This produces one fixed-dimensional representation per sensor modality despite differences in sampling rate and sequence length.

\paragraph{Multimodal Fusion}

After temporal aggregation, the 96-dimensional representations of the $M$ available modalities are concatenated into a $96M$-dimensional vector. A linear layer followed by layer normalization and a SiLU activation projects this vector into a shared 64-dimensional embedding, combining information extracted independently from the different sensing modalities. This embedding is the output of the backbone and forms the input to the subsequent classifier.

\paragraph{Distance-Based Classifier}

Instead of using a linear classification layer to directly produce class logits, we employ a distance-based classifier defined by class prototypes in the embedding space. Specifically, we learn one prototype $p_c\in\mathbb{R}^{64}$ for each class $c\in\{\textit{left},\textit{other},\textit{right}\}$. Given an embedding $z$ for an input window, the classifier computes its squared Euclidean distance to each class prototype and converts the corresponding negative distances into class probabilities using a softmax:

\begin{equation}
    \ell_c(z)=-\frac{\lVert z-p_c\rVert_2^2}{\tau},
    \qquad
    p(y=c\mid z)=\frac{\exp(\ell_c(z))}{\sum_k\exp(\ell_k(z))}.
    \label{eq:prototype_classifier}
\end{equation}

The positive global temperature $\tau$, shared across all classes, is learned jointly with the class prototypes and remaining model parameters. It controls the scale of the distance-based logits and thus the sharpness of the class probabilities, while leaving the nearest-prototype decision unchanged. The prototype-based classifier aligns the training objective with the user-adaptation procedure in \autoref{sec: user-adaptation}, as both classification and adaptation operate through class prototypes in the same Euclidean embedding space while keeping the backbone fixed during adaptation.

\paragraph{Training}
\label{sec: training}

For regularization, we apply dropout with a rate of $p=0.2$ to the fused embedding. As data augmentation, in addition to the 50\% window overlap described in \autoref{sec: preprocessing}, which increases the number of training samples through temporally shifted windows, we apply bilateral swapping with probability $p=0.5$. After standardization, swapping exchanges the normalized left- and right-ear sequences consistently across all modalities together with the labels \textit{left} and \textit{right}, while \textit{other} remains unchanged. This discourages reliance on accidental hardware- or ear-specific biases while preserving the ability to represent genuine bilateral asymmetries. All parameters are optimized end to end in PyTorch \cite{paszke2019pytorch} using cross-entropy loss and the Adam optimizer \cite{diederik2014adam} with a learning rate of $10^{-3}$ and a batch size of 256. Models are trained for at most 20 epochs, with 10\% of each training fold held out for validation under strict separation between training and validation data, and early stopping after three epochs without improvement in validation loss.

\paragraph{Deployment Footprint}
\label{sec:deployment}

Depending on the sensing modality, the resulting models contain 82.3--86.5k parameters, have checkpoint sizes of 330--359 KiB, and require 2.53--5.19 MMAC per 1 s bilateral input window. This small computational and memory footprint supports real-time execution on smartphones, as demonstrated in our cafeteria deployment (see \autoref{sec: mensa_user_study}), while remaining feasible on resource-constrained embedded platforms.

\subsection{User Adaptation}
\label{sec: user-adaptation}

Even for the same chewing activity, sensor signals can vary across users due to differences in anatomy and chewing behavior, earable placement, insertion depth, and ear-tip seal, as well as across foods (see \autoref{sec: selection_foods}). These factors can shift the embedding distribution of a new session relative to the source population. While model fine-tuning could compensate for such shifts, it requires backpropagation and sufficient calibration data to reliably update many parameters, making it less suitable for rapid personalization. We therefore adopt the lightweight, gradient-free Bayesian prototype adaptation principle introduced in BayaHAR \cite{burzer2026uncertainty}. Rather than updating the model backbone, adaptation uses a short labeled calibration session and closed-form maximum a posteriori (MAP) estimation to update the \textit{left} and \textit{right} class prototypes, while the \textit{other} prototype remains fixed.

\paragraph{Prototype Priors}

The prototype classifier learned during model training provides a single prototype $p_c$ for each class and is jointly optimized with the backbone, but does not characterize variability across users and sessions. We therefore use the frozen backbone to embed all training windows and estimate population-level prototype statistics for Bayesian adaptation. For each chewing class $c\in\{\textit{left},\textit{right}\}$, we compute a mean embedding for each subject--food combination. The mean and element-wise variance of these session-specific prototypes define the prior center $\mu_c^D$ and diagonal prior variance $v_c^D$, respectively, capturing between-session variability across users, wearing conditions, and foods. We additionally estimate the diagonal observation variance $v_c^{\mathrm{obs}}$ by pooling residuals of individual embeddings around their corresponding subject--food class mean, capturing within-session variability. For the \textit{other} class, which is not associated with a food condition, we average the class mean embeddings of individual source subjects to obtain the population prototype $\mu_{\textit{other}}^D$, which remains fixed during user adaptation. For inference without user-specific calibration, the population-level prototypes $\mu_c^D$ replace the prototypes $p_c$ learned during model training. For a target user $u$ and chewing class $c$, we define a Gaussian prior over the user-specific prototype $p_{u,c}$ and a Gaussian likelihood for an embedding $z_i$ as

\begin{equation}
    p(p_{u,c})
    =\mathcal{N}\!\left(p_{u,c}\mid\mu_c^D,
      \operatorname{diag}(v_c^D)\right),
    \qquad
    p(z_i\mid p_{u,c})
    =\mathcal{N}\!\left(z_i\mid p_{u,c},
      \operatorname{diag}(v_c^{\mathrm{obs}})\right).
    \label{eq:prototype_prior_model}
\end{equation}

\paragraph{MAP Prototype Update}

For a new user or session, we collect a short labeled calibration recording for each chewing-side and embed its windows using the frozen backbone. For each chewing class $c\in\{\textit{left},\textit{right}\}$, these embeddings form a support set $S_{u,c}$ of $N_c$ embeddings with mean $\mu_{u,c}^S$. Combining this user-specific evidence with the source-domain prior using Gaussian conjugacy yields the posterior variance and MAP prototype estimate,

\begin{equation}
    v_{u,c}^{\mathrm{post}}=\left((v_c^D)^{-1}
      +N_c(v_c^{\mathrm{obs}})^{-1}\right)^{-1}, 
      \qquad
    p_{u,c}^{\mathrm{MAP}}=v_{u,c}^{\mathrm{post}}\odot
      \left((v_c^D)^{-1}\odot\mu_c^D
      +N_c(v_c^{\mathrm{obs}})^{-1}\odot\mu_{u,c}^S\right),
\end{equation}

where all inverses and products are applied element-wise. The MAP prototype is a precision-weighted combination of the source-domain prior center and target-user support mean: with limited calibration data, it remains closer to the prior, while increasing user-specific evidence shifts it toward the target session. The resulting $p_{u,c}^{\mathrm{MAP}}$ replace $\mu_c^D$ for the left and right chewing classes, while $\mu_{\textit{other}}^D$ remains fixed. Adaptation therefore requires only support embedding and a closed-form prototype update, without fine-tuning the backbone.
\section{Evaluation}\label{sec: Evaluation}

We first describe the dataset in \autoref{sec: dataset} and evaluation protocol in \autoref{sec: eval protocol}. We then evaluate single sensing modalities in \autoref{sec: single sensors}, fused modalities in \autoref{sec: sensor fusion}, and robustness to acoustic interference in \autoref{sec: external validity_eval}. We subsequently assess how window-level predictions translate into meal-level chewing-side asymmetry estimates in \autoref{sec: asymmetry_bootstrap} and examine CHOMP's usability and acceptability in \autoref{sec: usability_acceptability}. We finally perform a cafeteria-situated evaluation of the deployed model in \autoref{sec: mensa_user_study}.

\subsection{Dataset}
\label{sec: dataset}

One participant was excluded due to corrupted synchronization files, resulting in a final dataset of 20 participants (13 male, 7 female). Each participant received a \texteuro15 voucher and the remaining contents of the bowl as compensation. Participants had a mean age of 26.3 years ($SD = 9.3$, $\text{range:}\ 20$--$55$), and all but three were right-handed. In accordance with the exclusion criterion of having no missing teeth, all included participants had a complete set of permanent teeth except for wisdom teeth, which were present in only three participants. Tooth restorations were reported by five participants. After preprocessing and segmentation into 1~s windows with 50\% overlap (see \autoref{sec: preprocessing}), the main dataset comprised 34,444 samples. The additional acoustic-interference recordings contributed 10,647 windows, yielding 45,091 windows when all recordings were included.

\subsection{Evaluation Protocol}
\label{sec: eval protocol}

We report macro precision, recall, and $F_1$ across the three classes \textit{left chewing}, \textit{right chewing}, and \textit{other}. Macro averaging gives equal weight to each class, preventing strong \textit{other}-class performance from obscuring weaker chewing-side recognition. Fold-level results are summarized using the median and interquartile range (IQR), which are less sensitive to outlying folds than the mean and standard deviation. Generalization to unseen subjects is evaluated using leave-one-subject-out (LOSO) cross-validation with 20 folds, while leave-one-food-out (LOFO) cross-validation assesses generalization to unseen foods. To prevent data leakage, all data-dependent training procedures, including feature standardization and training-window generation, use only the training portion of each fold, with no information from the held-out subject or food used during model training.

For LOSO, we evaluate the CHOMP base and adapted models using identical episodes for each held-out subject. Each episode corresponds to one food and simulates deployment in which a new user performs a short calibration before the model is applied to the remaining data from the same food. For each chewing side, the support set is constructed from 11~s of contiguous calibration data, yielding 20 overlapping 1~s windows. Query windows overlapping the support set are excluded, and the remaining query candidates comprise chewing windows from the same food together with \textit{other} windows from the held-out subject. We sample an equal number of query windows from each class so that the larger \textit{other} pool does not dominate the evaluation. The base model makes predictions without using the support labels, whereas the adapted model uses them for calibration. Both models are evaluated on exactly the same query windows to ensure direct comparability. Support selection and query sampling are repeated for ten random seeds per eligible food. For each held-out subject, each metric is first averaged across seeds within each food and then across eligible foods, yielding one score per subject. We report the median and IQR of these 20 subject-level scores. For LOFO, only the base model is evaluated on each held-out food's test fold, since adaptation to the held-out food would no longer represent generalization to an unseen food. Scores are summarized using the median and IQR across the held-out food folds.

\subsection{Single-Sensor Evaluation}
\label{sec: single sensors}

\begin{figure*}[t]
    \centering
    \includegraphics[width=\linewidth]{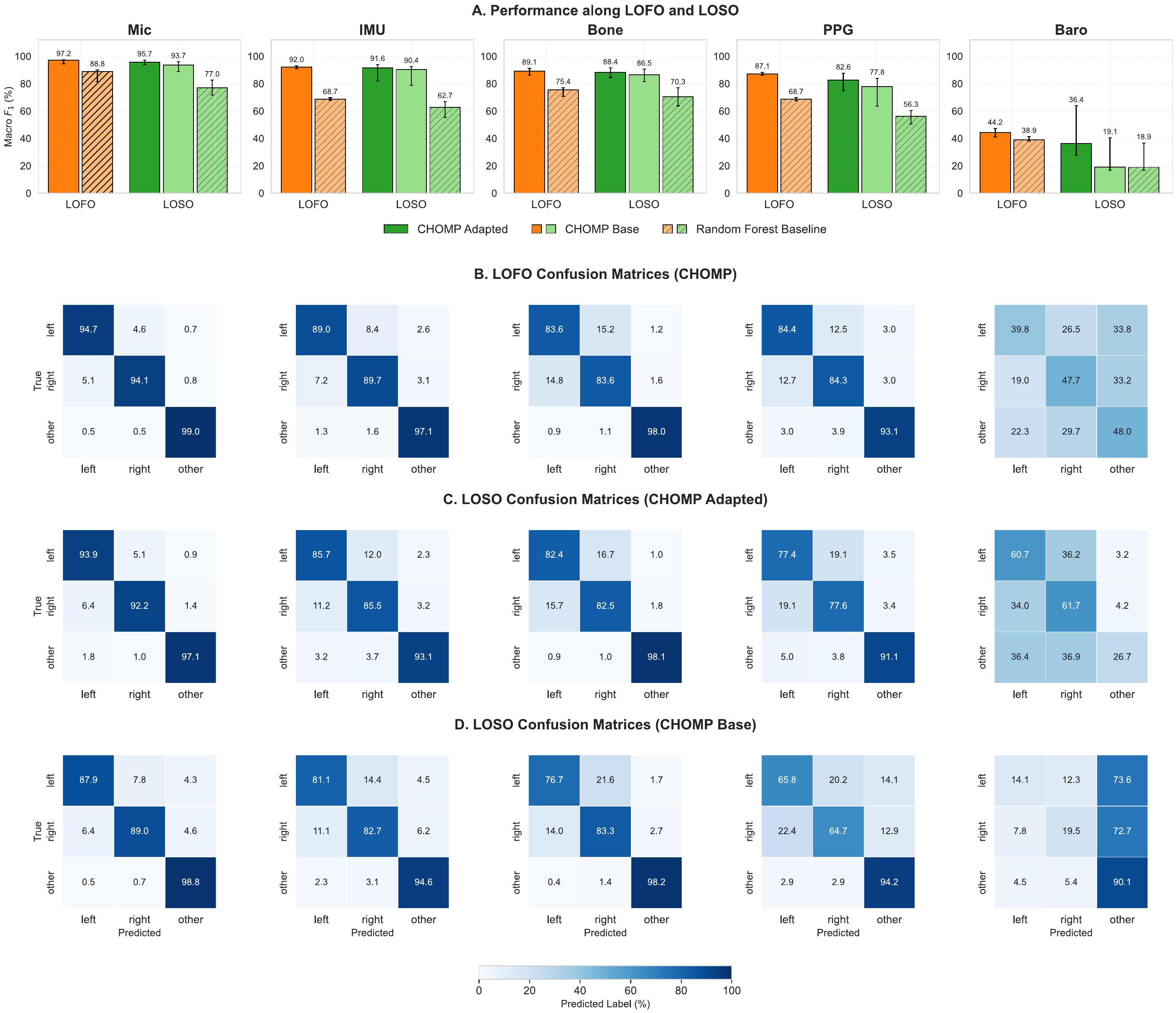}
    \caption{Single-sensor results. Classification performance is shown as median macro $F_1$ with interquartile ranges (Q1--Q3) for CHOMP and the Random Forest baseline under LOFO and LOSO evaluation. For LOSO, both the CHOMP base and user-adapted models are reported. The accompanying panels show row-normalized CHOMP confusion matrices for each sensing modality.}
    \Description{Comparison of CHOMP and Random Forest single-sensor performance under LOFO and LOSO evaluation, with row-normalized CHOMP confusion matrices.}
    \label{fig: Single Sensors}
\end{figure*}

We train separate CHOMP models for the microphones, bone-conduction microphone, IMU, PPG, and pressure sensor. As a conventional baseline motivated by prior chewing-detection work~\cite{lotfi_comparison_2020, ketmalasiri_imchew_2024}, we train a Random Forest~(RF) on the same standardized left- and right-ear feature sequences $L_m$ and $R_m$, aggregated using their mean and standard deviation. The RF uses 100 trees with a maximum depth of 20, at most 64 leaf nodes, at least five samples per leaf, square-root feature subsampling, bootstrap sampling, and balanced subsample class weights. These constraints provide an approximate size control relative to the corresponding single-sensor CHOMP models, although model complexity is not directly comparable between RFs and neural networks.

Results are shown in \autoref{tab: classification_results_single_sensors} and \autoref{fig: Single Sensors}. For the CHOMP base models, the microphones provide the strongest single-sensor performance, achieving a median macro $F_1$ of 97.2\% under LOFO and 93.7\% under LOSO, followed by the IMU at 92.0\% and 90.4\% and the bone-conduction microphone at 89.1\% and 86.5\%. PPG exhibits a more pronounced decrease from 87.1\% under LOFO to 77.8\% under LOSO, indicating greater sensitivity to inter-subject variation, while the pressure sensor performs substantially worse at 44.2\% and 19.1\%. CHOMP base models consistently outperform the RF baselines across all informative modalities and both evaluation protocols. On average, improvements amount to 13.8 percentage points under LOFO and 16.5 percentage points under LOSO, with the largest gains observed for the IMU.

\begin{figure*}[t]
    \centering
    \includegraphics[width=\linewidth]{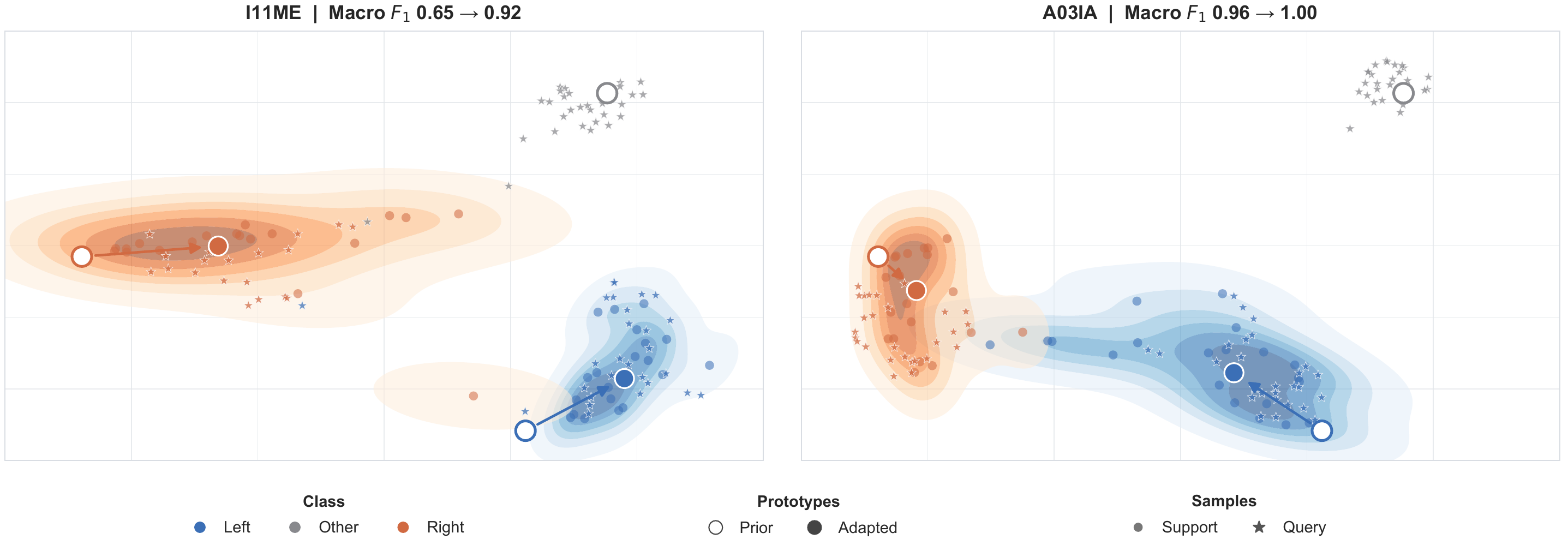}
    \caption{t-SNE visualization of user adaptation using microphone-only models on gummy bears. Colors indicate chewing class. Large open and filled circles show prior and adapted prototypes, respectively, with arrows indicating prototype updates through adaptation. Kernel density estimates (KDEs) show the distributions of the Left and Right support samples, while stars represent query samples. Panel titles report macro $F_1$ before and after adaptation.}
    \Description{}
    \label{fig: adaptation}
\end{figure*}

The confusion matrices show that most remaining errors concern the distinction between left- and right-sided chewing rather than chewing versus \textit{other}. For example, under LOSO, the bone-conduction model correctly identifies 98.2\% of \textit{other} windows, while 21.6\% of \textit{left} chewing windows are classified as \textit{right}. A short user-specific calibration session adapts the models and consistently improves chewing-side discrimination for unseen users. Averaged across the left and right chewing classes, correct classification between the sides increases from 88.5\% to 93.1\% for the microphones and from 81.9\% to 85.6\% for the IMU. Bone conduction shows the same pattern, increasing from 80.0\% to 82.5\%, while the strongest effect is observed for PPG, which improves from 65.3\% to 77.5\%. Consistent with these improvements in chewing-side discrimination, adaptation increases median macro $F_1$ from 93.7\% to 95.7\% for the microphones, 90.4\% to 91.6\% for the IMU, 86.5\% to 88.4\% for bone conduction, and 77.8\% to 82.6\% for PPG. User adaptation is visualized in \autoref{fig: adaptation}, showing how support samples collected during calibration guide prototype updates, resulting in better alignment with the query samples and decision boundaries tailored to individual users and sessions.

% These results demonstrate the benefit of modeling the temporal evolution of paired-ear feature sequences over fixed summary statistics, while user-specific adaptation further improves chewing-side recognition across unseen users.

% The full baseline specification is provided in \autoref{app: random_forest}.

\subsection{Multimodal Fusion Evaluation}\label{sec: sensor fusion}

\begin{figure*}[t]
    \centering
    \includegraphics[width=\linewidth]{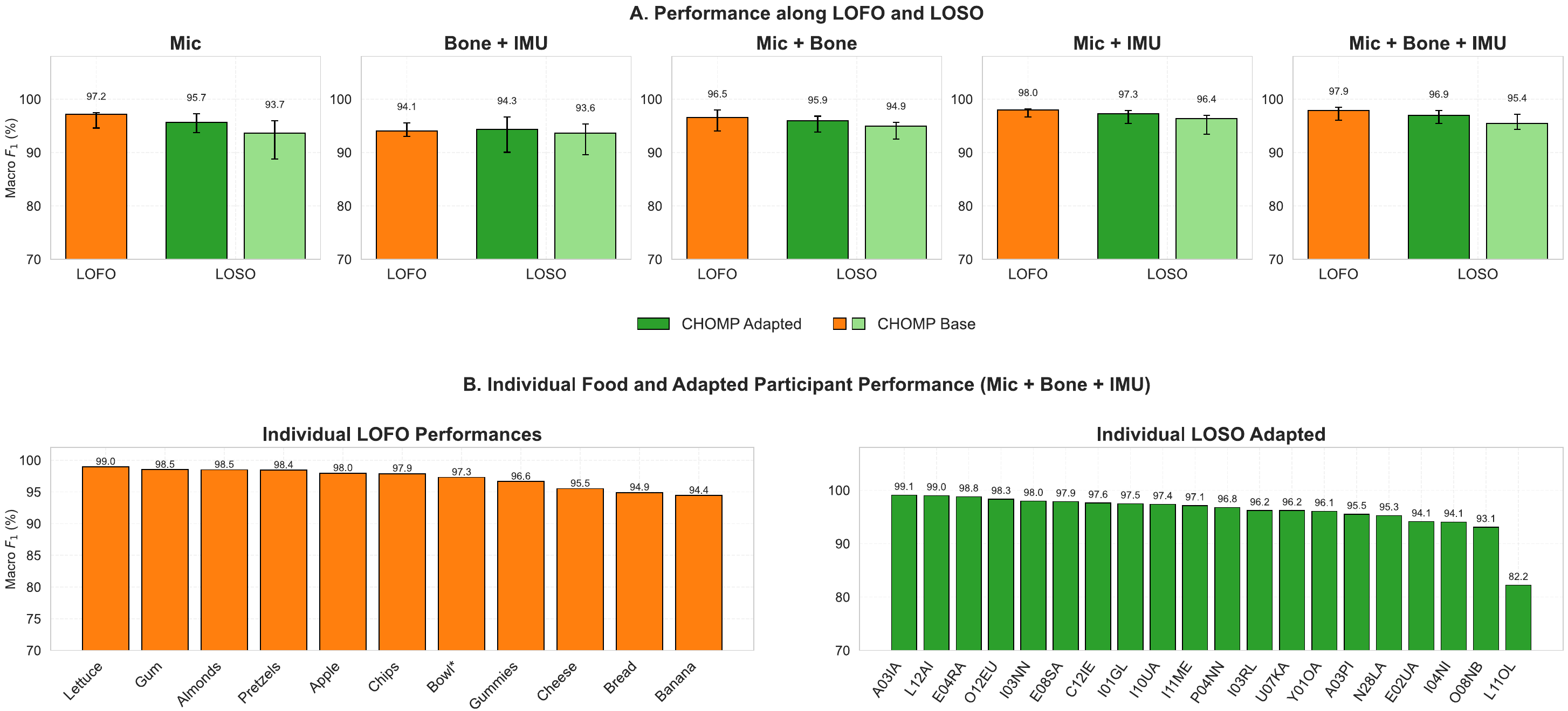}
    \caption{Sensor-fusion results. Median macro $F_1$ with interquartile ranges (Q1--Q3) is shown for combinations of microphones, bone-conduction microphone, and IMU under LOFO and LOSO evaluation. Microphone-only performance is included as a reference. The lower panels show performance for each held-out food and adapted participant using all three modalities. The bowl result uses reduced training data because bowl runs are three times more frequent than individual food recordings.}
    \Description{Comparison of microphone, bone-conduction microphone, and IMU fusion configurations under LOFO and LOSO evaluation. The lower panels show individual held-out food performance and adapted held-out participant performance for the full three-modality model.}
    \label{fig: Multi Sensors}
\end{figure*}

Next, we evaluated whether multimodal fusion could further improve chewing-side detection, following the fusion architecture described in \autoref{sec: model_architecture}. Based on the preceding single-sensor analysis, we selected the microphones, bone-conduction microphone, and IMU, which showed the strongest individual performance. We evaluated all pairwise combinations and their three-modality fusion, resulting in four configurations.

Results are detailed in \autoref{tab: classification_results_fusion_sensors} and \autoref{fig: Multi Sensors}. Among the two-sensor combinations, microphones and IMU achieve the highest performance under both LOFO (98.0\%) and LOSO (96.4\%), improving over the microphones-only benchmark of 97.2\% and 93.7\%, respectively. The full configuration of all three modalities performs similarly at 97.9\% under LOFO and 95.4\% under LOSO, while the bone-conduction microphone and IMU reach 94.1\% and 93.6\%, respectively. User adaptation improves all fusion configurations under LOSO, increasing median macro $F_1$ from 93.6\% to 94.3\% for Bone+IMU, from 96.4\% to 97.3\% for Mic+IMU, from 94.9\% to 95.9\% for Mic+Bone, and from 95.4\% to 96.9\% for the full three-modality configuration. Mic+IMU therefore achieves the highest adapted LOSO performance at 97.3\%. Examining the full three-modality model across held-out foods, performance ranges from 94.4\% for banana to 99.0\% for lettuce, and under adapted LOSO, 19 of 20 participants achieve at least 93.1\% macro $F_1$.

\subsection{Performance under Acoustic Interference}
\label{sec: external validity_eval}

To assess how acoustic interference affects the different sensing modalities, we use the additional semi-hard cheese and snack-pretzel sessions recorded under three noise conditions: dining-hall ambience and construction noise played through an external speaker, and music played directly through the earphones (see \autoref{sec: external_validity}). Rather than testing generalization from clean training data to unseen noise, this analysis evaluates performance when the corresponding acoustic conditions are represented during training. We use the adapted CHOMP model and restrict both adaptation and evaluation to cheese and pretzel recordings. Each noise condition uses recordings collected under that condition, while the No Noise setting follows the same procedure using clean recordings of the same foods as a reference. We additionally evaluate an All Noises setting combining the six noisy food variants. For each setting, LOSO training is repeated with the corresponding sessions included, while the held-out subject remains unseen during training and adaptation uses only recordings from the evaluated condition. This condition-matched setup allows us to compare how acoustic interference affects the sensing modalities and their combinations.

\begin{figure*}[t]
    \centering
    \includegraphics[width=\linewidth]{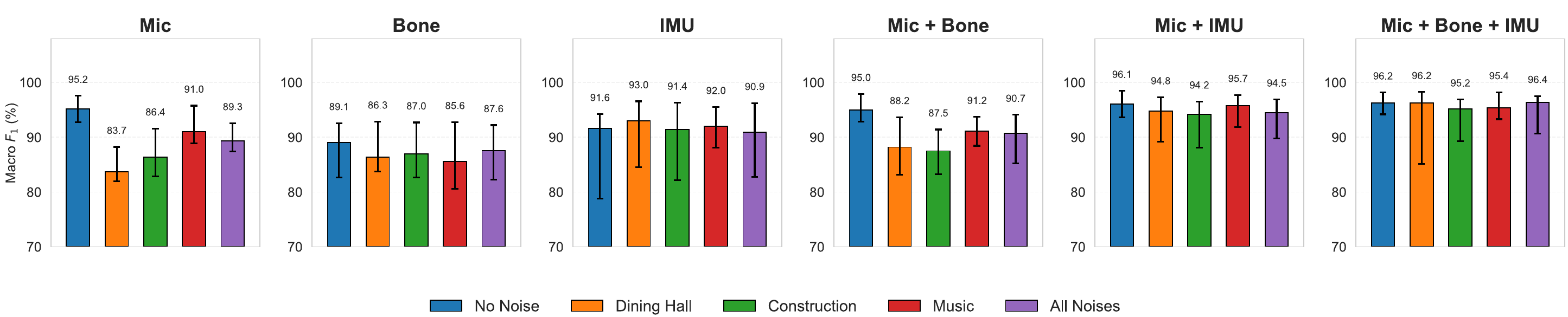}
    \caption{Robustness to acoustic interference for the adapted CHOMP model under LOSO. Median macro $F_1$ with interquartile ranges (Q1--Q3) is shown for clean recordings, dining-hall noise, construction noise, earphone music, and all noise conditions combined. Adaptation and evaluation are restricted to the corresponding cheese and pretzel recordings, with clean recordings of the same foods serving as the No Noise reference.}
    \Description{LOSO performance of the adapted CHOMP model using single and fused sensor configurations on cheese and pretzel recordings under no noise, dining-hall noise, construction noise, earphone music, and all noise conditions combined.}
    \label{fig: Audio Robust Evaluation}
\end{figure*}

Results are shown in \autoref{tab: classification_results_audio_robust} and \autoref{fig: Audio Robust Evaluation}. The clean reference highlights a clear difference between acoustic and motion sensing under the tested conditions. The microphone model decreases from 95.2\% median macro $F_1$ without noise to 83.7\% under dining-hall noise and 86.4\% under construction noise, while earphone music has a smaller effect at 91.0\%. In contrast, the IMU remains close to its 91.6\% clean performance, achieving between 90.9\% and 93.0\% across the noise settings. Bone conduction also varies less than the microphones, although at a lower overall level, ranging from 85.6\% to 87.6\% compared with 89.1\% without noise. Multimodal fusion maintains performance closer to the clean reference. Mic+IMU achieves 96.1\% without noise and between 94.2\% and 95.7\% across the individual noise conditions, while the full Mic+Bone+IMU configuration achieves 96.2\% without noise and between 95.2\% and 96.2\% across the individual conditions. When all noises are combined, the two configurations reach 94.5\% and 96.4\%, respectively. Overall, acoustic interference affects microphone-only classification more strongly than IMU-based classification, while fusion with motion sensing maintains performance close to the clean reference across the tested conditions.

\afterpage{
\clearpage
\begin{table*}[t]
\centering
\scriptsize
\caption{Single-sensor classification performance for CHOMP and the Random Forest (RF) baseline under LOFO and LOSO evaluation (percent). Macro $F_1$ is reported as median (Q1–Q3), while precision and recall are reported as medians. For LOSO, both the CHOMP base and user-adapted models are reported.}
\newcolumntype{F}{>{\centering\arraybackslash}p{2.55cm}}
\newcolumntype{M}{>{\centering\arraybackslash}p{0.75cm}}
\begin{tabular*}{\textwidth}{@{\extracolsep{\fill}}p{2cm} p{2cm} F M M F M M@{}}
\toprule
& & \multicolumn{3}{c}{\textbf{LOFO}} & \multicolumn{3}{c}{\textbf{LOSO}} \\
\cmidrule(lr){3-5} \cmidrule(lr){6-8}
\textbf{Sensor} & \textbf{Model}
& \textbf{Macro $F_1$ (IQR)} & \textbf{Pre.} & \textbf{Rec.}
& \textbf{Macro $F_1$ (IQR)} & \textbf{Pre.} & \textbf{Rec.} \\
\midrule
Mic & CHOMP Adapted & -- & -- & -- & \textbf{95.7 (93.8-97.3)} & \textbf{96.2} & \textbf{95.7} \\
 & CHOMP Base & \textbf{97.2 (94.6-97.5)} & \textbf{97.2} & \textbf{97.1} & 93.7 (88.8-96.0) & 94.5 & 93.9 \\
 & RF Baseline & 88.8 (81.3-90.0) & 89.1 & 88.8 & 77.0 (71.7-82.6) & 80.0 & 77.6 \\
\addlinespace[3pt]
IMU & CHOMP Adapted & -- & -- & -- & \textbf{91.6 (82.1-93.9)} & \textbf{92.6} & \textbf{91.7} \\
 & CHOMP Base & \textbf{92.0 (91.0-93.1)} & \textbf{92.1} & \textbf{91.9} & 90.4 (78.7-92.5) & 91.9 & 90.5 \\
 & RF Baseline & 68.7 (67.7-69.7) & 68.7 & 68.7 & 62.7 (55.3-66.9) & 63.8 & 62.6 \\
\addlinespace[3pt]
Bone & CHOMP Adapted & -- & -- & -- & \textbf{88.4 (84.5-91.5)} & \textbf{89.3} & \textbf{88.4} \\
 & CHOMP Base & \textbf{89.1 (86.1-91.3)} & \textbf{88.9} & \textbf{89.4} & 86.5 (81.4-90.5) & 88.7 & 86.9 \\
 & RF Baseline & 75.4 (70.8-77.1) & 74.8 & 76.3 & 70.3 (63.7-76.9) & 70.0 & 71.2 \\
\addlinespace[3pt]
PPG & CHOMP Adapted & -- & -- & -- & \textbf{82.6 (74.8-87.7)} & \textbf{84.5} & \textbf{82.9} \\
 & CHOMP Base & \textbf{87.1 (86.1-88.3)} & \textbf{87.2} & \textbf{86.9} & 77.8 (63.6-83.9) & 79.2 & 78.4 \\
 & RF Baseline & 68.7 (67.4-69.7) & 68.6 & 68.7 & 56.3 (50.6-60.4) & 58.7 & 59.0 \\
\addlinespace[3pt]
Baro & CHOMP Adapted & -- & -- & -- & \textbf{36.4 (27.7-63.8)} & \textbf{39.4} & \textbf{42.8} \\
 & CHOMP Base & \textbf{44.2 (41.0-47.3)} & \textbf{55.8} & \textbf{46.8} & 19.1 (16.8-40.5) & 21.3 & 33.7 \\
 & RF Baseline & 38.9 (38.2-41.3) & 49.6 & 43.7 & 18.9 (16.6-36.7) & 31.9 & 33.4 \\
\addlinespace[3pt]
\bottomrule
\end{tabular*}
\label{tab: classification_results_single_sensors}
\end{table*}

\begin{table*}[t]
\centering
\scriptsize
\caption{Sensor-fusion classification performance for different combinations of microphones, bone-conduction microphone, and IMU under LOFO and LOSO evaluation (percent). Macro $F_1$ is reported as median (Q1–Q3), while precision and recall are reported as medians. For LOSO, both the CHOMP base and user-adapted models are reported.}
\newcolumntype{F}{>{\centering\arraybackslash}p{2.55cm}}
\newcolumntype{M}{>{\centering\arraybackslash}p{0.75cm}}
\begin{tabular*}{\textwidth}{@{\extracolsep{\fill}}p{2cm} p{2cm} F M M F M M@{}}
\toprule
& & \multicolumn{3}{c}{\textbf{LOFO}} & \multicolumn{3}{c}{\textbf{LOSO}} \\
\cmidrule(lr){3-5} \cmidrule(lr){6-8}
\textbf{Sensor Units} & \textbf{Model}
& \textbf{Macro $F_1$ (IQR)} & \textbf{Pre.} & \textbf{Rec.}
& \textbf{Macro $F_1$ (IQR)} & \textbf{Pre.} & \textbf{Rec.} \\
\midrule
Bone + IMU & CHOMP Adapted & -- & -- & -- & \textbf{94.3 (90.0-96.7)} & \textbf{94.9} & \textbf{94.3} \\
 & CHOMP Base & \textbf{94.1 (93.0-95.6)} & \textbf{94.2} & \textbf{94.0} & 93.6 (89.6-95.3) & 94.3 & 93.6 \\
\addlinespace[3pt]
Mic + IMU & CHOMP Adapted & -- & -- & -- & \textbf{97.3 (95.5-97.9)} & \textbf{97.5} & \textbf{97.3} \\
 & CHOMP Base & \textbf{98.0 (96.7-98.2)} & \textbf{97.9} & \textbf{98.0} & 96.4 (93.5-97.0) & 96.8 & 96.4 \\
\addlinespace[3pt]
Mic + Bone & CHOMP Adapted & -- & -- & -- & \textbf{95.9 (93.8-96.8)} & \textbf{96.4} & \textbf{96.0} \\
 & CHOMP Base & \textbf{96.5 (94.0-97.9)} & \textbf{96.5} & \textbf{96.5} & 94.9 (92.5-95.7) & 95.8 & 95.0 \\
\addlinespace[3pt]
Mic + Bone + IMU & CHOMP Adapted & -- & -- & -- & \textbf{96.9 (95.5-97.9)} & \textbf{97.2} & \textbf{97.0} \\
 & CHOMP Base & \textbf{97.9 (96.1-98.5)} & \textbf{97.9} & \textbf{97.9} & 95.4 (94.3-97.2) & 96.2 & 95.5 \\
\addlinespace[3pt]
\bottomrule
\end{tabular*}
\label{tab: classification_results_fusion_sensors}
\end{table*}

\begin{table*}[t]
\centering
\scriptsize
\caption{Acoustic-interference performance of the user-adapted CHOMP models under LOSO evaluation for semi-hard cheese and snack pretzel recordings (percent). No Noise denotes the corresponding clean recordings, while All Noises combines all noisy food variants. Macro $F_1$ is reported as median (Q1–Q3) across held-out participants.}
\newcolumntype{F}{>{\centering\arraybackslash}p{2.25cm}}
\begin{tabular*}{\textwidth}{@{\extracolsep{\fill}}p{2.2cm} F F F F F@{}}
\toprule
\textbf{Sensor Units}
& \textbf{No Noise}
& \textbf{Dining Hall}
& \textbf{Construction}
& \textbf{Earphone Music}
& \textbf{All Noises} \\
\midrule
Mic
& \textbf{95.2 (92.7-97.6)}
& 83.7 (81.9-88.2)
& 86.4 (82.9-91.6)
& 91.0 (88.9-95.8)
& 89.3 (87.4-92.6) \\
\addlinespace[2pt]

Bone
& 89.1 (82.7-92.6)
& 86.3 (83.7-92.9)
& 87.0 (82.6-92.7)
& 85.6 (80.6-92.8)
& 87.6 (82.2-92.2) \\
\addlinespace[2pt]

IMU
& 91.6 (78.8-94.2)
& \textbf{93.0 (84.5-96.6)}
& \textbf{91.4 (82.2-96.3)}
& \textbf{92.0 (88.1-95.5)}
& \textbf{90.9 (82.7-96.2)} \\
\addlinespace[2pt]

\midrule

Mic + IMU
& 96.1 (93.6-98.5)
& 94.8 (89.2-97.3)
& 94.2 (88.1-96.5)
& \textbf{95.7 (91.9-97.7)}
& 94.5 (89.8-96.9) \\
\addlinespace[2pt]

Mic + Bone
& 95.0 (92.9-97.9)
& 88.2 (83.2-93.7)
& 87.5 (83.2-91.4)
& 91.2 (88.4-93.7)
& 90.7 (85.3-94.1) \\
\addlinespace[2pt]

Mic + Bone + IMU
& \textbf{96.2 (94.2-98.2)}
& \textbf{96.2 (85.1-98.3)}
& \textbf{95.2 (89.3-96.9)}
& 95.4 (93.3-98.2)
& \textbf{96.4 (90.7-97.5)} \\
\bottomrule
\end{tabular*}
\label{tab: classification_results_audio_robust}
\end{table*}
\clearpage
}

\subsection{Meal-Level Detection of Chewing-Side Preference}
\label{sec: asymmetry_bootstrap}

As the window-level results do not reveal how much chewing must be observed before a chewing-side preference becomes detectable, we aggregated held-out LOSO predictions after user adaptation into synthetic meals corresponding to 1, 5, and 10 minutes of chewing. With one non-overlapping 1\,s window per second, these durations comprise $N = 60$, $300$, and $600$ consecutive windows. Consecutive predictions are not independent, however. Within the recorded chewing sequences, classification errors were clustered with a design effect of $d = 1.72$ \cite{kish1965survey}, meaning that $N$ consecutive windows carry the evidence of only $N/d$ independent ones. Because the windows of a synthetic meal are drawn independently, we composed each meal of $N/d$ windows, i.e., 34, 174, and 348 windows after rounding to the nearest even number. We varied the meal composition from 52/48 to 80/20 in steps of 2\%, rounded to whole windows. For each participant and composition, we generated 5,000 meals, half with each side dominant, by sampling with replacement from that participant's windows of each chewing side. All sensor configurations were evaluated on identical meals. A meal was classified as asymmetric if a two-sided exact binomial test against equal preference ($H_0: p_{\mathrm{right}} = 0.5$, $\alpha = 0.05$) on the windows predicted as left or right was significant, and the asymmetry was counted as detected if the predicted dominant side matched the true one.

\begin{figure}[t]
    \centering
    \includegraphics[width=\linewidth]{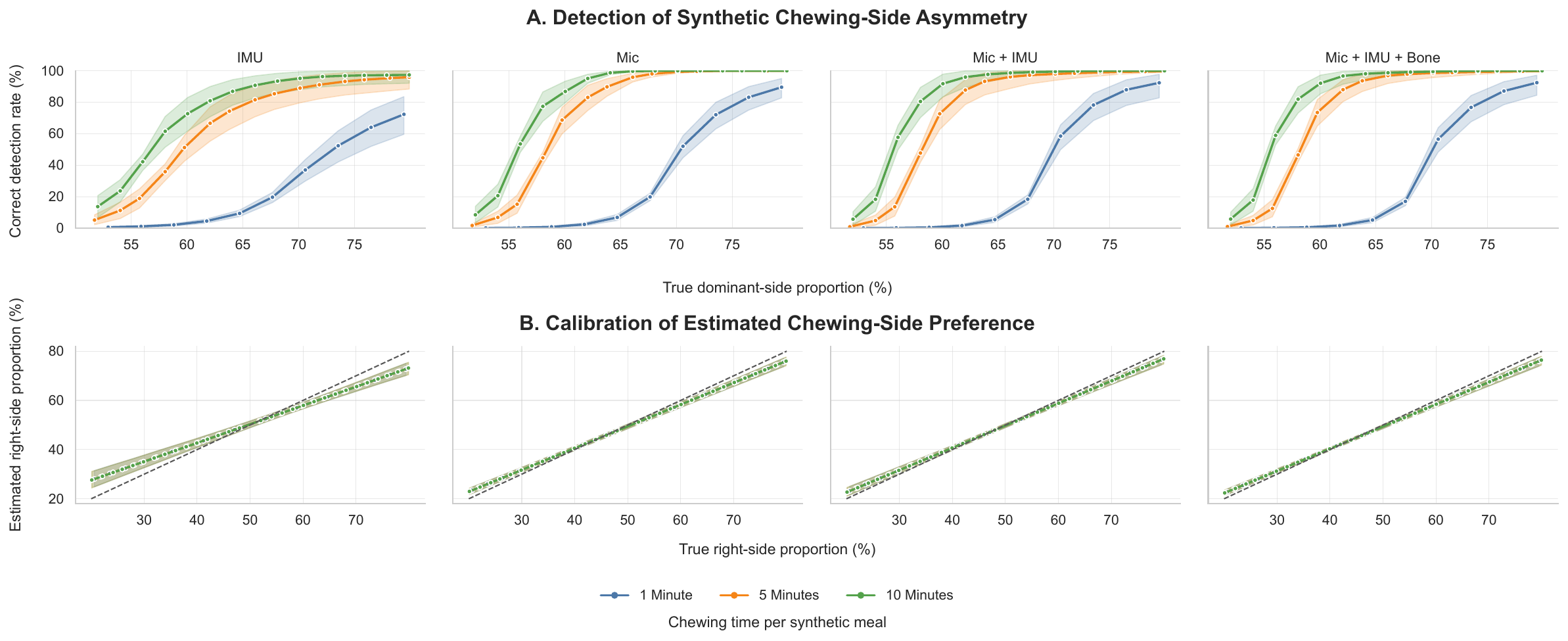}
    \caption{Detection and calibration of chewing-side asymmetry in synthetic meals constructed from adapted LOSO predictions. Meals of 1, 5, and 10 minutes were simulated with 34, 174, and 348 effectively independent 1\,s windows, accounting for clustered classification errors ($d = 1.72$). A. Rate of correct-side detection. B. Estimation versus true right-side proportion (dashed: perfect calibration). Lines show means across the 20 held-out participants, and shaded bands show 95\% bootstrap confidence intervals.}
    \Description{A two-row figure with four columns: IMU, microphone, microphone plus IMU, and microphone plus IMU plus bone conduction. Each panel has three curves for synthetic meals of 1, 5, and 10 minutes of chewing, with shaded confidence bands. In the top row, the rate of correct asymmetry detection rises with the true dominant-side proportion, shown from 52 to 80 percent, and with meal duration. For the three microphone-based configurations, the 10 and 5 minute curves exceed 95 percent from dominant-side proportions of about 62 and 66 percent, whereas the 1 minute curves stay below 10 percent up to about 65 percent, rise steeply around 70 percent, and reach about 90 percent at 80 percent. The IMU curves are lower throughout: its 10 and 5 minute curves reach 95 percent at about 70 and 78 percent, and its 1 minute curve reaches about 70 percent. In the bottom row, the estimated right-side proportion is plotted against the true proportion from 20 to 80 percent, with a dashed perfect-calibration diagonal. The curves for all three durations lie on top of each other and are slightly flatter than the diagonal: true proportions of 20 and 80 percent are estimated as about 23 and 76 percent by the microphone-based configurations and as about 28 and 73 percent by the IMU.}
    \label{fig:asymmetry_bootstrap}
\end{figure}

With the microphone and IMU combined, a 70/30 preference was detected in at least 80\% of meals after approximately 1.6 minutes of chewing, and a 60/40 preference after approximately 6.1 minutes (\autoref{fig:asymmetry_bootstrap}). Adding the bone-conduction sensor yielded nearly identical results (1.6 and 5.9 minutes). The microphone alone required 1.7 and 7.1 minutes, respectively, whereas the IMU alone required 2.8 minutes for a 70/30 preference and did not reach the 80\% level for a 60/40 preference within ten minutes. For comparison, error-free window classification would require 0.6 and 1.8 minutes for the same task. The estimated share followed the true share but was attenuated toward 50\%. A true share of 80\%, for example, was estimated as 77.2\% with the microphone and IMU, 77.1\% with the bone-conduction sensor added, 76.6\% with the microphone alone, and 72.8\% with the IMU alone.

\subsection{Usability and Acceptability Analysis}\label{sec: usability_acceptability}

\begin{figure*}[t]
    \centering
    \includegraphics[width=\linewidth]{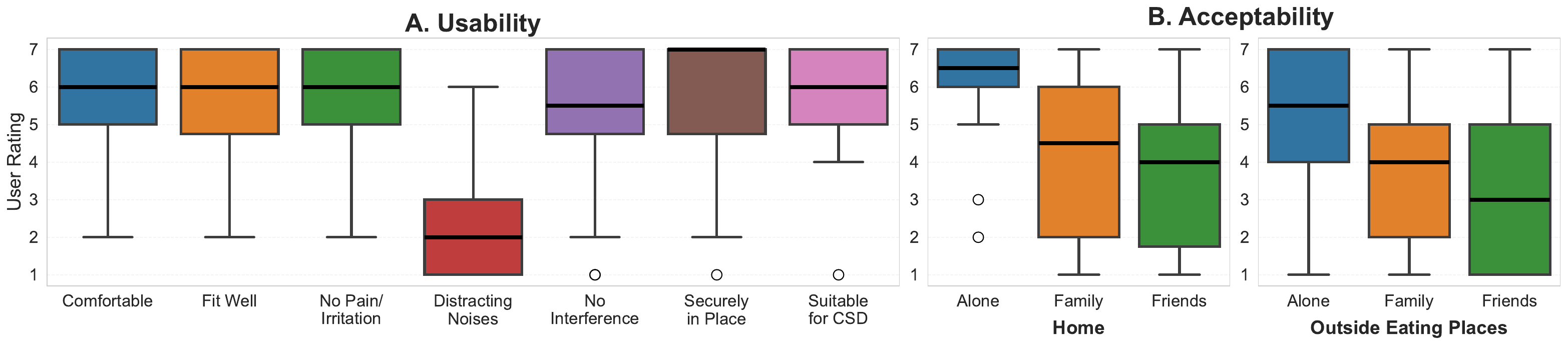}
    \caption{Usability and acceptability of OpenEarable 2.0 for chewing-side detection. A. Participant ratings of usability and comfort during the study. B. Acceptability ratings across eating situations and social contexts. Public eating places comprise dining hall, fast-food place, café, and restaurant.}
    \Description{Boxplots of the usability and acceptability questionnaire responses on seven-point scales.}
    \label{fig: Usability Eval}
\end{figure*}

All items evaluating the usability and acceptability of using OpenEarable 2.0 for chewing-side detection were answered on a 7-point Likert scale (\textit{1 = Totally Disagree}, \textit{7 = Totally Agree}, or \textit{1 = Very Uncomfortable}, \textit{7 = Very Comfortable}, respectively). Overall results are summarized in \autoref{fig: Usability Eval}, and the full questionnaires are provided in \autoref{app: post_quest}.

\paragraph{Usability}
Results indicate that OpenEarable 2.0 devices were generally comfortable to wear ($M = 5.60$) and, on average, fit the ears well ($M = 5.55$). Participants reported little to no pain or irritation ($M = 5.80$). Movements of the earables caused only few distracting noises ($M=2.35$, inverse item) and interfered only slightly with eating ($M = 5.15$). Participants predominantly indicated that the earables stayed securely in place while moving or chewing ($M = 5.65$) and considered them suitable for detecting the chewing-side ($M = 5.65$).

\paragraph{Acceptability}
Participants showed a clear preference for using earables for chewing-side detection when eating alone ($M = 5.38$) compared with eating with family ($M = 3.75$) or friends ($M = 3.40$). Moreover, using the earables at home ($M = 4.68$) was preferred over use in public eating environments, including dining halls ($M = 4.30$), fast-food places ($M = 4.12$), cafes ($M = 3.97$), and restaurants ($M = 3.82$). Accordingly, deploying earables for chewing-side detection when eating at home alone was most favorable ($M = 6.05$).

\subsection{Cafeteria-Situated User Study}\label{sec: mensa_user_study}

\begin{table*}[b]
    \centering
    \footnotesize
    \caption{Seven-point Likert items used in the cafeteria study to assess calibration ease, perceived prediction correctness, overall perceived accuracy, and system suitability. All items were rated from 1 = Totally disagree to 7 = Totally agree, with higher ratings indicating a more favorable assessment.}
    \label{tab:cafeteria-questionnaire-items}
    \begin{tabular}{lp{11.5cm}}
    \toprule
    \textbf{Criterion} & \textbf{Item} \\
    \midrule
    Calibration & The short left/right calibration was easy to complete. \\

    \addlinespace

    Side Prediction & While I was chewing, CHOMP generally displayed the side on which I felt I was chewing. \\

    \addlinespace

    Other Prediction & While I was not chewing, CHOMP generally displayed \textit{Other}. \\

    \addlinespace

    Perceived Accuracy & CHOMP's predictions felt accurate during this meal. \\

    \addlinespace

    System Suitability & Based on this experience, I feel that CHOMP is suitable for predicting my chewing-side during meals. \\
    \bottomrule
    \end{tabular}
\end{table*}

\begin{figure*}[t]
    \centering

    \begin{subfigure}[t]{0.42\textwidth}
        \centering
        \caption{Selection of Cafeteria Dishes.}
        \label{fig: mensa_meals}
        \includegraphics[width=\linewidth]{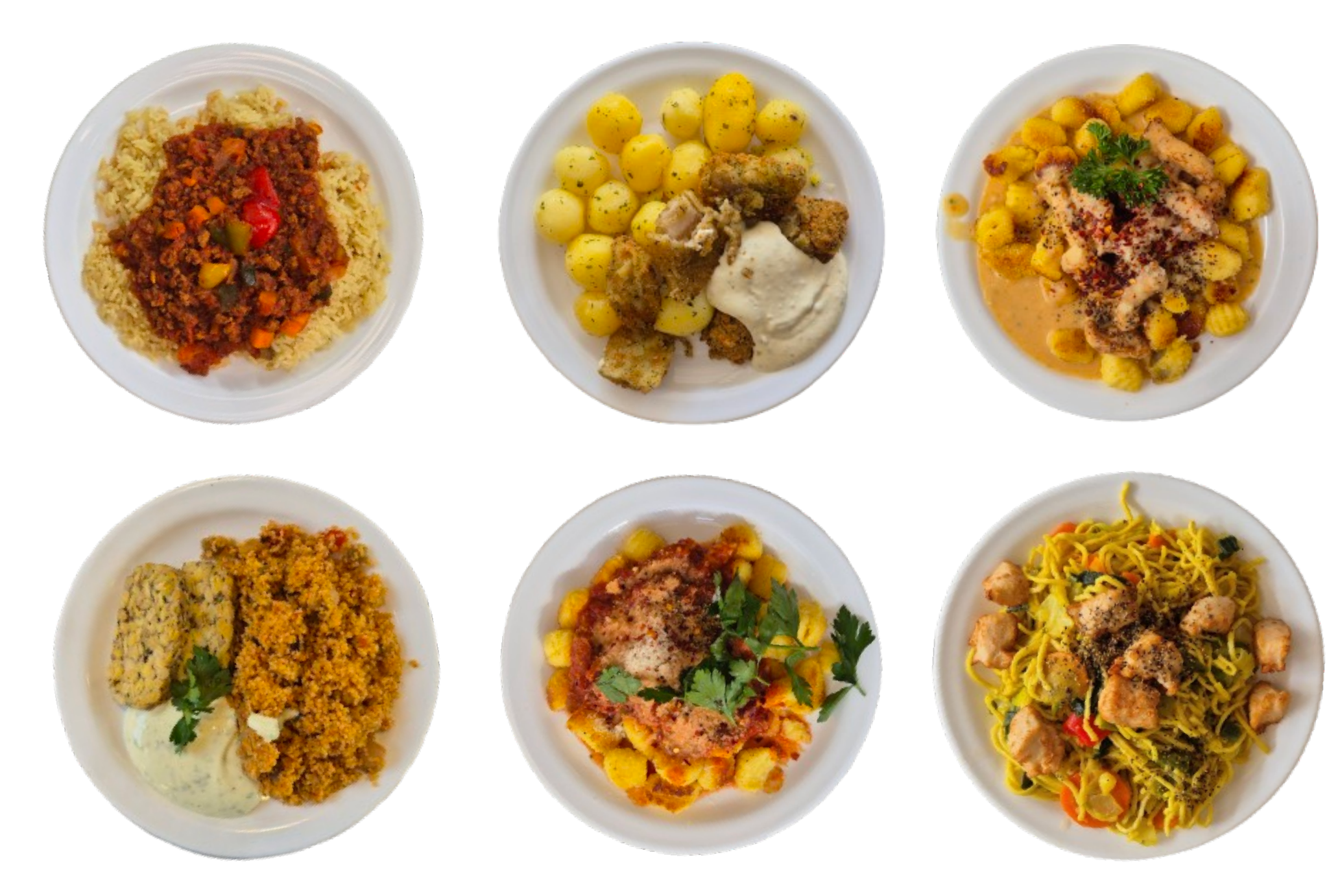}
    \end{subfigure}
    \hfill
    \begin{subfigure}[t]{0.54\textwidth}
        \centering
        \caption{User Study Results.}
        \label{fig: mensa_user_study_boxplot}
        \includegraphics[width=\linewidth]{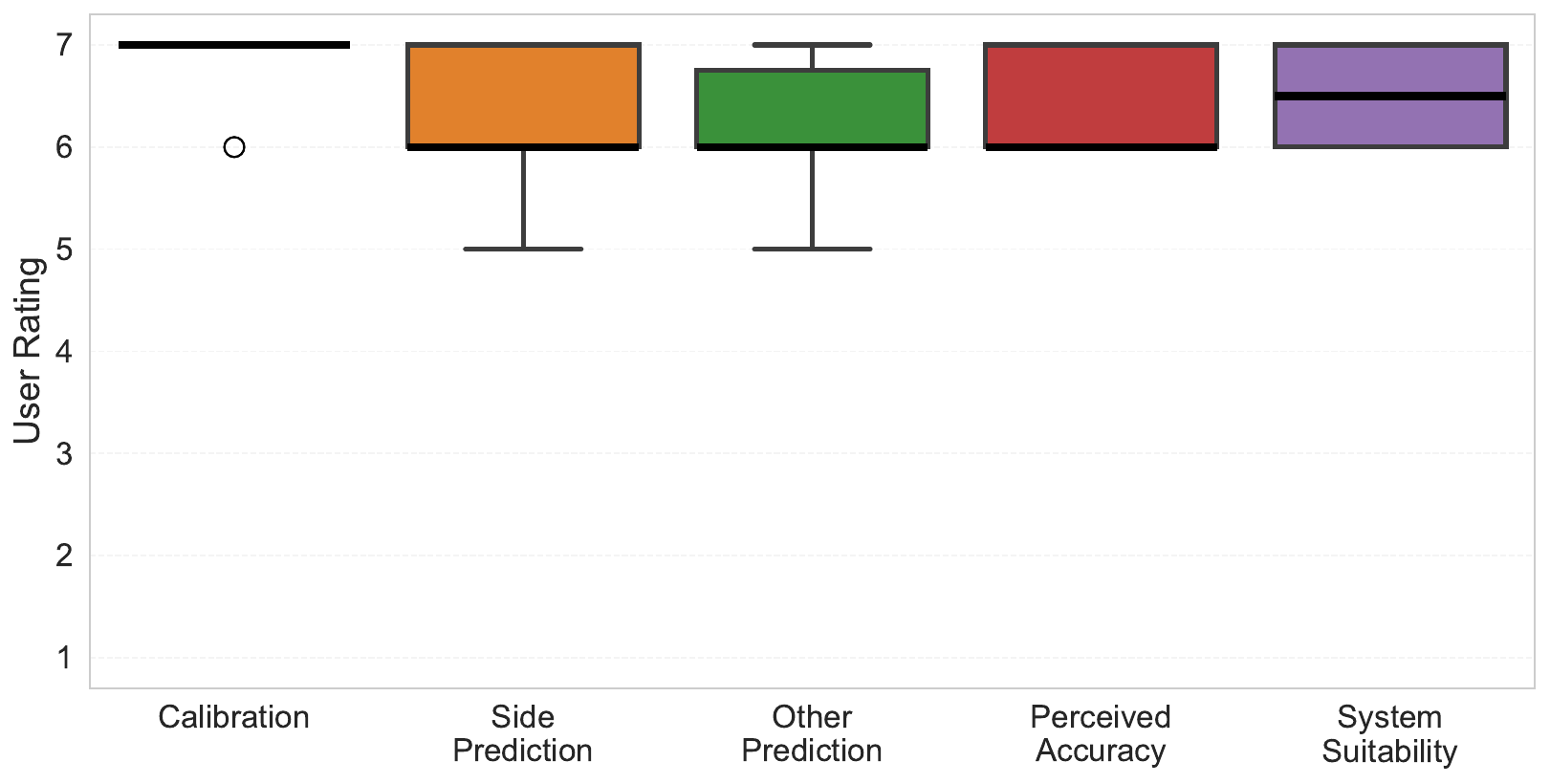}
    \end{subfigure}

    \caption{Cafeteria-situated user study. A. Example dishes selected by participants from the regular cafeteria menu. B. Participant ratings of calibration ease, side-prediction correctness, other-prediction correctness, perceived accuracy, and system suitability on a 7-point Likert scale.}
    \Description{Cafeteria-deployed user study.; A. Example dishes selected by participants from the regular cafeteria menu, including for example gnocchi or bami goreng with chicken; B. Cafeteria-situated user study participant ratings as a boxplot (see \autoref{sec: mensa_user_study} text).}
    \label{fig: mensa_double_fig}
\end{figure*}

To evaluate CHOMP beyond the controlled laboratory setting, we deployed a real-time version of it during regular meals in the cafeteria of the Karlsruhe Institute for Technology. Due to bandwidth constraints of the live streaming setup, the deployed system used the IMU-only model, which reached a median macro $F_1$ of 91.6\% under LOSO for the adapted model (\autoref{sec: single sensors}). While CHOMP primarily targets offline analysis of completed meals, the live deployment allowed participants to compare its predictions with their own perception while eating. As obtaining window-level ground truth during unconstrained meals would substantially interfere with natural eating, we assessed perceived rather than measured performance~\cite{bedri_fitnibble_2022}. We recruited ten participants (6 male, 4 female; $M_{\mathrm{age}} = 23.2$, $SD = 3.5$) on site among diners eating alone. None had participated in the main data collection (\autoref{sec: dataset}), making all participants unseen users. Each received a \texteuro10 voucher. After informed consent and a short pre-questionnaire, participants were fitted with the OpenEarable 2.0 devices. For calibration, they chewed one bite exclusively on each side for about 10~s. The resulting 20 windows per side were used to adapt CHOMP's \textit{left} and \textit{right} class prototypes, while the \textit{other} prototype was retained from training. Participants then ate a dish of their choice from the regular cafeteria menu (\autoref{fig: mensa_meals}) without instructions regarding chewing-side. CHOMP classified the most recent paired 1~s window every 0.5~s and displayed \textit{left}, \textit{right}, or \textit{other} on a smartphone placed in front of the participant, with a latency of about 1~s. Participants were asked to keep the display in view while otherwise eating naturally. Monitoring lasted 10.0~min on average ($SD = 4.3$). After the meal, participants rated five study-specific items on a 7-point Likert scale (\textit{1 = Totally Disagree}, \textit{7 = Totally Agree}; \autoref{tab:cafeteria-questionnaire-items}). Ratings were high across all items (\autoref{fig: mensa_user_study_boxplot}). Participants found the calibration easy to complete ($M = 6.80$), reported that CHOMP displayed \textit{other} when they were not chewing ($M = 6.10$), and that its side prediction generally matched the side on which they felt they were chewing ($M = 6.20$). They also perceived CHOMP's predictions as accurate during the meal ($M = 6.40$) and considered it suitable for predicting their chewing-side during meals ($M = 6.50$). Among windows predicted as \textit{left} or \textit{right}, the participant-wise dominant predicted side accounted for $M = 60.0\%$ of windows ($SD = 6.5\%$).

% all new participants that were not part of the training data set

% several different menus (also show in the picture)

% only people eating alone as they should concentrate on the application

% for calibration, protocol that left/right is selected in parallel

% instruct them to eat normally

% do the calibration properly otherwise does not work well

% tell them clearly that this has a 1s/half second delay 

% report left/right results in the end, at least descriptively

% report as means with boxplots (like usability) or generally change to medians and iqr througout (then also in user study)

% FitNibble: https://dl.acm.org/doi/10.1145/3490099.3511154

% EarBit: https://dl.acm.org/doi/10.1145/3130902

% Potentially Interesting: How accurate does it feel: https://dl.acm.org/doi/10.1145/3491102.3501915

% Potential Soundbite: CHOMP was additionally deployed end-to-end during unrestricted cafeteria meals. Because continuous ground-truth labeling would interfere with natural eating, this deployment was not used to estimate classification accuracy. Instead, it assessed practical operation and participants' perceived correctness of the resulting predictions under an in-situ eating condition.

% During evaluation, participant or food separation is preserved according to the respective protocol. After model selection, the deployment backbone is retrained on all available subjects and windows for a fixed number of epochs equal to the median validation-selected epoch across the leave-one-subject-out folds.
\section{Discussion}\label{sec: Discussion}

In this work, we introduce CHOMP, the first system for chewing-side detection using earphones. Microphones provide the strongest single-sensor performance, reaching median macro $F_1$ scores of 97.2\% under LOFO and 95.7\% under LOSO evaluation, while multimodal fusion further improves performance to 98.0\% under LOFO and 97.3\% after user adaptation under LOSO. The system also maintains high performance under three types of acoustic interference. We demonstrate that aggregating held-out predictions enables meal-level estimation of chewing-side asymmetry and show that users rated CHOMP as highly usable and particularly acceptable when eating alone. Finally, we deployed the IMU-only model live during regular cafeteria meals, where previously unseen users perceived its predictions as accurate and suitable for chewing-side monitoring. We next explore these findings in greater detail (\autoref{sec: exploration of results}), discuss their real-world applicability and implications for longitudinal chewing-side monitoring (\autoref{sec: implications}), and identify remaining limitations and directions for future work (\autoref{sec: limitations}).

\subsection{Exploration of Results}\label{sec: exploration of results}

\subsubsection{Earphones for Chewing-Side Detection}

This study is the first to demonstrate the feasibility of employing earphones for chewing side detection. Moreover, we show that this task can be reliably addressed using TWS earphones, now the dominant consumer form factor, which inherently limits cross-correlative features due to network-induced jitter at short temporal scales \cite{hunn_introducing_2024}. CHOMP requires sufficient alignment to construct paired 1~s windows, but does not depend on such sample-level inter-ear features, showing that the bilateral information available within aligned windows is sufficient for accurate chewing-side recognition. CHOMP also compares favorably with previous wearable approaches. Earlier systems using neck clips, headbands, or glasses reported accuracies or $F_1$ scores below 90\%~\cite{chung_glasses-type_2017, nakamura_automatic_2021, wang_wearable_2021}, whereas CHOMP already exceeds this level using microphones or IMU alone under both LOFO and LOSO evaluation. CHOMP therefore extends prior evidence substantially by demonstrating high performance across participants, mechanically diverse foods, multiple sensing modalities, and more challenging evaluation conditions. The results further suggest that this performance depends not only on sensor placement, but also on how the bilateral signals are modeled. In contrast to the RF baseline, CHOMP retains temporal order through modality-specific bidirectional GRUs and weighted temporal aggregation. Its consistent advantage across informative modalities indicates that the temporal evolution within each chewing window contains discriminative information beyond fixed summary statistics. Multimodal fusion provides an additional benefit, with microphones and IMU forming the strongest combination. Adding bone conduction does not further improve median performance under controlled conditions, while still making performance more robust against acoustic interference. The modality results also provide guidance for future earable implementations. Microphones are the strongest standalone modality, while the IMU performs well independently and provides the clearest complementary benefit in fusion. Bone conduction performs somewhat lower but, together with the IMU, still enables strong chewing-side recognition without relying on air-conduction microphones. PPG also carries side-related information, but its larger LOSO drop and stronger benefit from user adaptation suggest greater sensitivity to individual physiology or placement. Pressure sensing performs substantially worse despite its demonstrated utility for chewing detection~\cite{hossain_ear_2023}. This may reflect the dependence of ear-canal pressure sensing on a consistent seal, which specialized devices can provide more reliably than a general-purpose earable~\cite{roddiger_earrumble_2021, ando_canalsense_2017}. Overall, these results establish TWS earphones as a suitable platform for chewing-side detection while highlighting clear trade-offs between sensing modalities and their combinations.

\begin{table*}[t]
\centering
\footnotesize
\caption{Correlations of macro $F_1$ with usability questionnaire items under LOSO (mean adapted episode score per participant), and between macro $F_1$ and log-transformed food firmness under LOFO (held-out food score). Significant correlations are marked with *. Statistical power is limited by 20 participants, 10 foods, and ceiling effects in the questionnaire items.}
\label{tab:correlations}
    \begin{tabular}{
        p{3.2cm}
        >{\centering\arraybackslash}p{2.0cm}
        >{\centering\arraybackslash}p{1.8cm}
        >{\centering\arraybackslash}p{2.4cm}
        >{\centering\arraybackslash}p{2.0cm}}
\toprule
\textbf{Sensor Units} & \textbf{Comfortable} & \textbf{Fit Well} & \textbf{Securely in Place} & \textbf{Firmness (log)} \\
\midrule
Mic & 0.24 & 0.27 & 0.60\rlap{$^{*}$} & -0.08 \\
Bone & 0.29 & -0.03 & 0.39 & 0.03 \\
IMU & 0.31 & 0.13 & 0.43 & -0.19 \\
\midrule
Bone + IMU & 0.39 & 0.21 & 0.51\rlap{$^{*}$} & -0.09 \\
Mic + Bone & 0.32 & -0.03 & 0.32 & -0.06 \\
Mic + IMU & 0.41 & 0.21 & 0.53\rlap{$^{*}$} & -0.13 \\
Mic + Bone + IMU & 0.37 & 0.17 & 0.50\rlap{$^{*}$} & -0.07 \\
\bottomrule
\end{tabular}
\end{table*}

\subsubsection{Inter-Subject and Inter-Food Variability}

Differences between participants remain more pronounced than food-related variation for several sensing modalities and can arise from differences in anatomy, earable placement, ear-tip seal, and individual chewing behavior. The short user-specific calibration addresses part of this variability by adapting the left and right class prototypes while keeping the feature extractor fixed. Adaptation improves all informative single-sensor and fusion configurations, and for the full multimodal configuration, 19 of 20 unseen participants achieve at least 93.1\% macro $F_1$. This suggests that substantial user-specific variation can be accommodated through lightweight prototype adaptation, although it is not eliminated entirely. The correlations in \autoref{tab:correlations} further suggest a possible role of earable attachment. Secure placement shows the most consistent association with performance, with positive correlations across all configurations and significant correlations for microphones and three fusion configurations. Comfort is also positively associated with performance throughout, whereas self-reported fit shows only weak associations. Given the small sample and questionnaire ceiling effects, these relationships should be interpreted as exploratory rather than causal. Nevertheless, practical systems could incorporate fit or seal checks~\cite{demirel_unobtrusive_2024, kuttner_earresp-ans_2026}, while personalized ear tips may further reduce placement variability~\cite{zhao_fit2ear_2024, amesaka_flexear-tips_2025}. In contrast, performance varies little across the food set. The full multimodal configuration achieves $94.4 - 99.0\%$ across held-out foods, while correlations with log-firmness remain weak across all sensor configurations ($r=-0.19$ to $0.03$; \autoref{tab:correlations}). Thus, measured firmness alone does not explain the remaining performance differences. 

\subsubsection{Robustness Beyond Controlled Conditions}

Robustness to acoustic interference is particularly important for an earphone-based system as its best-performing modality, the microphones, must coexist with audio playback and ambient sound during everyday use. Previous audio-based approaches to chewing-side detection have not evaluated such interference~\cite{nakamura_automatic_2020a,nakamura_automatic_2020b,nakamura_automatic_2021}, while robustness to environmental interference has generally remained unexplored in prior wearable approaches~\cite{yamasakiObjectiveAssessmentActual2015, wang_wearable_2021, chung_glasses-type_2017, kim_earchew_2022}. Our results show that high performance is maintained under the evaluated acoustic conditions, while the IMU and bone-conduction microphone provide complementary sensing pathways that are less directly exposed to airborne sound. This is important because everyday meals are unlikely to occur under consistently controlled acoustic conditions. The cafeteria deployment extends this perspective from the sensing environment to the eating context itself. Previously unseen users consumed regular meals without prescribed chewing sides while a live IMU-only version of CHOMP performed user-adapted inference. Although the study assesses perceived rather than objective classification accuracy~\cite{bedri_fitnibble_2022}, the positive user ratings indicate that the complete workflow of short calibration, sensing, and inference can operate during naturalistic meals.

\subsection{Real-World Applicability and Clinical Implications}
\label{sec: implications}

Considering that approximately one-third of the global population is affected by TMDs~\cite{alqutaibi_global_2025, zielinski_meta-analysis_2024}, unobtrusive solutions for longitudinal monitoring of everyday jaw function could provide substantial value. As described in \autoref{sec: rw_csp_tmd}, current TMD assessment and diagnosis largely rely on self-reports and clinical examinations~\cite{diernberger_self-reported_2008, national_academies_of_sciences_prevalence_2020}, providing only episodic insight into natural eating behavior and its development over time. CHOMP addresses part of this gap by enabling repeated measurements of chewing-side behavior during meals. In particular, our meal-level analysis shows how window-level predictions can be aggregated into estimates of chewing-side asymmetry. These estimates could complement existing clinical instruments, such as symptom diaries, by providing an objective record of how chewing behavior changes across meals and over time.

This is particularly relevant because CSP has been associated with TMDs~\cite{lopez-cedrun_jaw_2017, yap_sleeping_2024} and may itself contribute to their development~\cite{barcellos_prevalence_2011, tiwari_chewing_2017}, while also being associated with other long-term alterations such as changes in craniofacial structures~\cite{jiang_analysis_2015}. Earphones offer an attractive form factor for obtaining such measurements because they are routinely worn and can capture chewing-related signals without requiring an additional dedicated wearable. Our cafeteria deployment further demonstrates that a lightweight IMU-only configuration can operate during regular meals and that previously unseen users perceived its predictions as accurate and suitable for chewing-side monitoring. Together with the strong performance across held-out foods and participants, as well as under acoustic-interference conditions, these findings support the feasibility of extending chewing-side sensing beyond controlled laboratory recordings.

The intended use of CHOMP is primarily retrospective rather than based on continuous real-time feedback. A user could record one or more meals and subsequently review aggregated chewing-side behavior, reducing the need to attend to the system while eating. In this role, CHOMP could provide clinicians and patients with longitudinal behavioral information that is difficult to obtain through episodic clinical examinations alone.

\subsection{Limitations and Future Work}\label{sec: limitations}

Despite the promising implications discussed above, several limitations bound the conclusions of this work. First, the primary evaluation used a controlled protocol in which participants ate while seated and chewed exclusively on one side, providing reliable ground truth but not capturing spontaneous side switching or other behavioral variability during natural meals. The cafeteria study extends the evaluation to a more naturalistic setting, but assesses perceived rather than measured performance because continuous window-level ground truth would interfere with natural eating.

Second, we did not collect labels for simultaneous bilateral chewing because doing so would have substantially increased data collection effort. This follows prior chewing-side work (see \autoref{sec: rw_detecting_csp_and_tmd}) and the tendency for individual chews to have a working side~\cite{iida_kinematics_2023, turcio_does_2016}, but prevents CHOMP from explicitly representing bilateral or otherwise ambiguous chewing. Future models could therefore include a bilateral or uncertain state.

Third, all participants in the main study had complete permanent dentition except for wisdom teeth, limiting generalization to older adults and individuals with missing teeth, dental prostheses, or other conditions affecting mastication. CHOMP has also not been evaluated in a clinical TMD population or as a diagnostic instrument. Longitudinal studies with individuals diagnosed with TMD are therefore needed to determine how CHOMP-derived measures relate to symptoms, disease progression, and clinically meaningful outcomes.

Finally, while CHOMP enables repeated measurement of chewing-side behavior, we have not investigated how this information could support behavioral change. The live feedback interface demonstrates the feasibility of presenting chewing-side predictions during a meal, but whether feedback can modify chewing behavior and how it should be designed to produce lasting effects remains to be studied.
\section{Conclusion}\label{sec: Conclusion}

In this paper, we presented CHOMP, the first system for chewing-side detection using earphones. Our multimodal sensor fusion approach achieves median macro $F_1$ scores of 98.0\% under LOFO and 97.3\% after user adaptation under LOSO evaluation, representing the highest classification performance reported to date for chewing-side detection by wearable systems, while maintaining high performance across diverse food types and acoustic-interference conditions. Beyond window-level classification, we demonstrate that aggregating predictions enables meal-level estimation of chewing-side asymmetry. We further deployed an IMU-only version of CHOMP during regular cafeteria meals, where previously unseen users perceived its predictions as accurate and suitable for chewing-side monitoring. By enabling repeated and unobtrusive measurement of chewing-side behavior, CHOMP provides a pathway toward longitudinal assessment of jaw function in everyday life. Such information could complement existing clinical assessments by providing insights into potentially maladaptive chewing-side asymmetries and their development over time.

\begin{acks}
The authors thank Ms. Katharina Schiesser for excellent technical assistance during the food texture measurements.
This work was funded by the Deutsche Forschungsgemeinschaft (DFG, German Research Foundation) --  GRK2739/2 -- Project Nr. 447089431 -- Research Training Group: KD2School –- Designing Biosignal-Adaptive Systems for Decision-Making Processes.

\end{acks}

\bibliographystyle{ACM-Reference-Format}
\bibliography{sample-base}

\newpage

\appendix

\section{Appendix: Pre-Questionnaire}\label{app: pre_quest}

The pre-questionnaire collected basic demographic and dental information from all participants. Note that the dental diagram referenced in the questionnaire is not included here due to copyright restrictions.\newline

\begin{description}
    \item[Gender:] Please select your gender.  
    \newline \textit{Response options:} Male; Female; Diverse; Prefer not to say.

    \par\addvspace{0.5em}

    \item[Age:] Please enter your age in years.  
    \newline \textit{Response options:} Open Answer.

    \par\addvspace{0.5em}

    \item[Handedness:] Are you left-handed or right-handed?  
    \newline \textit{Response options:} Left-handed; Right-handed.

    \par\addvspace{0.5em}

    \item[Missing Teeth:] Are you missing any teeth that have not been replaced with replacement materials or implants (i.e., do you have gaps in your teeth)? Missing wisdom teeth are excluded!  
    \newline \textit{Response options:} Yes; No.

    Note: Participants were excluded from the study if they answered \textit{Yes}.

    \par\addvspace{0.5em}

    \item[Wisdom Teeth:] Do you have your wisdom teeth?  
    \newline \textit{Response options:} Yes; Yes, but not erupted; No; Partially or Partially erupted (Please mark accordingly in the diagram on the accompanying paper sheet).

    \par\addvspace{0.5em}

    \item[Teeth Repairments:] Do you have fillings, crowns, implants, or other dental restorations? If yes, please indicate the affected teeth in the diagram on the accompanying paper sheet and specify the material used for each tooth (if known).  
    \newline \textit{Response options:} Yes (Please mark accordingly in the diagram); No.

    \par\addvspace{0.5em}

    \item[Dental Malocclusions:] Do you have any other dental malocclusions (e.g., overbite or underbite)? Please list them and briefly describe if necessary.  
    \newline \textit{Response options:} Open Answer.
\end{description}
\newpage

\section{Appendix: Post-Questionnaire}\label{app: post_quest}

We reviewed the existing literature for questionnaires assessing the usability and acceptability of wearing devices while eating, but identified only a single relevant example \cite{liu_intelligent_2012}. This served as an initial reference point, although the final questionnaire was developed specifically for the present study. \newline

\par\addvspace{1.0em}
\noindent\textbf{Usability}

\par\addvspace{0.5em}

\noindent All items were answered on a 7-point Likert scale  
(\textit{1 = Totally Disagree}, \textit{7 = Totally Agree}).

\par\addvspace{0.5em}

\noindent\textbf{Instructions:} Please answer the questions on the usability and comfort of the chewing detection with earables below.

\par\addvspace{0.5em}
\begin{enumerate}
    \item The earables were comfortable to wear throughout the study.
    \item The earables fit my ears well.
    \item Wearing the earables did not cause any pain or irritation.
    \item Movements of the earables during chewing created distracting noises in my ears.
    \item The earables did not interfere with my eating.
    \item The earables stayed securely in place while I moved or chewed.
    \item I felt that the earables were suitable for chewing-side detection.
    \item Do you have any additional comments or suggestions regarding the comfort or usability of the earables for chewing-side detection? (\textit{Open Answer})
\end{enumerate}

\par\addvspace{1.0em}
\noindent\textbf{Acceptability}

\par\addvspace{0.5em}

\noindent All items were answered on a 7-point Likert scale  
(\textit{1 = Very Uncomfortable}, \textit{7 = Very Comfortable}).

\par\addvspace{0.5em}

\noindent\textbf{Instructions:} Please answer the questions on wearing the earables for chewing-side detection in different situations.

\noindent\textbf{Leading Question:} I would feel comfortable wearing the earables for chewing-side detection in the following places and situations:

\par\addvspace{0.5em}
\begin{enumerate}
    \item At home (alone)
    \item At home (with family)
    \item At home (with friends)
    \item In the dining hall (alone)
    \item In the dining hall (with family)
    \item In the dining hall (with friends)
    \item At a fast-food place (alone)
    \item At a fast-food place (with family)
    \item At a fast-food place (with friends)
    \item In a café (alone)
    \item In a café (with family)
    \item In a café (with friends)
    \item In a restaurant (alone)
    \item In a restaurant (with family)
    \item In a restaurant (with friends)
    \item Do you have any additional comments or suggestions about the comfort of wearing the earables during chewing-side detection in different situations? (\textit{Open Answer})
\end{enumerate}

\newpage

\section{Appendix: Bowl Characteristics}\label{app: bowl_characteristics}

For the study, a bowl was ordered from Dean and David\footnote{\href{https://deananddavid.com/}{Official Website}}. To align with the study design, a \textit{Mix Your Own} bowl was selected, allowing participants to configure their meal within predefined options. This ensured both diversity in composition and overall comparability across participants. The available choices were as follows:  

\begin{enumerate}
    \item Base: salad and rice (jasmine fragrant rice OR quinoa wholegrain rice)
    \item Crunchy topping: croutons OR crunchy onions OR peanuts OR walnuts OR sunflower seeds
    \item Hard vegetable: cucumber OR carrot OR bell pepper
    \item Soft vegetable: mango OR beetroot OR tomato
    \item Protein: chicken OR beef OR salmon OR halloumi OR vegetarian chicken OR falafel
    \item Dressing: choice of participant
\end{enumerate}

\end{document}